\documentclass[pre,aps,showpacs,byrevtex,amsmath,amssymb,twocolumn]{revtex4}
\usepackage{graphicx}

\begin{document}

\title{Modeling DNA Structure, Elasticity and Deformations at the Base-pair Level}

\author{Boris Mergell}
\email{mergell@mpip-mainz.mpg.de}
\author{Mohammad R. Ejtehadi}
\author{Ralf Everaers}
\email{everaers@mpipks-dresden.mpg.de}

\affiliation{Max-Planck-Institut f\"{u}r Polymerforschung,
 Postfach 3148, D-55021 Mainz,
 Germany}

\date{\today}

\begin{abstract}
  We present a generic model for DNA at the base-pair level. We use a
  variant of the Gay-Berne potential to represent the stacking energy
  between neighboring base-pairs. The sugar-phosphate backbones are
  taken into account by semi-rigid harmonic springs with a non-zero
  spring length. The competition of these two interactions and the
  introduction of a simple geometrical constraint leads to a stacked
  right-handed B-DNA-like conformation. The mapping of the presented
  model to the Marko-Siggia and the Stack-of-Plates model enables us
  to optimize the free model parameters so as to reproduce the
  experimentally known observables such as persistence lengths, mean
  and mean squared base-pair step parameters. For the optimized model
  parameters we measured the critical force where the transition from
  B- to S-DNA occurs to be approximately $140\mbox{pN}$. We observe an
  overstretched S-DNA conformation with highly inclined bases that
  partially preserves the stacking of successive base-pairs.
\end{abstract}

\pacs{87.14.Gg,87.15.Aa,87.15.La,61.41.+e}

\maketitle

\section{Introduction}

Following the discovery of the double helix by Watson and
Crick~\cite{WatsonCrick_nat_53}, the structure and elasticity of DNA
has been investigated on various length scales.  X-ray diffraction
studies of single crystals of DNA oligomers have led to a detailed
picture of possible DNA
conformations~\cite{Dickerson_sci_82,DickersonFromAToZ} with atomistic
resolution. Information on the behavior of DNA on larger scales is
accessible through NMR~\cite{James_methenz_95} and various optical
methods~\cite{Millar_jcp_82,Schurr_arevpc_86},
video~\cite{Perkins_sci_94} and electron
microscopy~\cite{Boles_jmb_90}. An interesting development of the last
decade are nanomechanical experiments with {\em individual} DNA
molecules~\cite{Smith_sci_92,Smith_sci_96,Cluzel_sci_96,Heslot_pnas_97,Allemand_pnas_98}
which, for example, reveal the intricate interplay of supercoiling on
large length scales and local denaturation of the double-helical
structure.

Experimental results are usually rationalized in the framework of two
types of models: base-pair steps and variants of the continuum elastic
worm-like chain.  The first, more local, approach describes the
relative location and orientation of neighboring base pairs in terms
of intuitive parameters such as twist, rise, slide, roll
etc.~\cite{CalladineDrew_jmb_84,Dickerson_emboj_89,LuOlson_jmb_99,OlsonNomenclature_jmb_01}.
In particular, it provides a mechanical interpretation of the
biological function of particular sequences~\cite{CalladineDrew99}.
The second approach models DNA on length scales beyond the helical
pitch as a worm-like chain (WLC) with empirical parameters describing
the resistance to bending, twisting and
stretching~\cite{MarkoSiggia_mm_94,MarkoSiggia_mm_95}. The results are
in remarkable agreement with the nanomechanical experiments mentioned
above~\cite{Perkins_sci_95}.  WLC models are commonly used in order to
address biologically important phenomena such as
supercoiling~\cite{Cozzarelli_90,SchlickOlson_jmb_92,ChiricoLangowski_bp_94}
or the wrapping of DNA around histones~\cite{Schiessel_prl_01}. In
principle, the two descriptions of DNA are linked by a systematic
coarse-graining procedure: From given (average) values of rise, twist,
slide etc. one can reconstruct the shape of the corresponding helix
on large
scales~\cite{CalladineDrew_jmb_84,ElHassanCalladine_londa_97,CalladineDrew99}.
Similarly, the elastic constant characterizing the continuum model are
related to the local elastic energies in a stack-of-plates
model~\cite{Hern_epjb_98}.

Difficulties are encountered in situations which cannot be described
by a linear response analysis around the undisturbed (B-DNA) ground
state.  This situation arises routinely during cellular processes and
is therefore of considerable biological
interest~\cite{CalladineDrew99}.  A characteristic feature, observed
in many nanomechanical experiments, is the occurrence of plateaus in
force-elongation
curves~\cite{Smith_sci_96,Cluzel_sci_96,Allemand_pnas_98}. These
plateaus are interpreted as structural transitions between
microscopically distinct states.  While atomistic simulations have
played an important role in identifying possible local structures such
as S- and P-DNA~\cite{Cluzel_sci_96,Allemand_pnas_98}, this approach
is limited to relatively short DNA segments containing several dozen
base pairs. The behavior of longer chains is interpreted on the basis
of stack-of-plates models with step-type dependent parameters and free
energy penalties for non-B steps.  Realistic force-elongation are
obtained by a suitable choice of parameters and as the consequence of
constraints for the total extension and twist (or their conjugate
forces)~\cite{ChatenayMarko_pre_01}.  Similar models describing the
non-linear response of B-DNA to stretching~\cite{HZhou_prl_99} or
untwisting~\cite{Barbi_pl_99,Cocco_prl_99} predict stability
thresholds for B-DNA due to a combination of more realistic,
short-range interaction potentials for rise with twist-rise coupling
enforced by the sugar-phosphate backbones.

Clearly, the agreement with experimental data will increase with the
amount of details which is faithfully represented in a DNA model.
However, there is strong evidence both from atomistic
simulations~\cite{Lavery_Meso_bpj_99} as well as from the analysis of
oligomer crystal structures~\cite{ElHassanCalladine_ptrs_97} that the
base-pair level provides a sensible compromise between conceptual
simplicity, computational cost and degree of reality.  While Lavery et
al.~\cite{Lavery_Meso_bpj_99} have shown that the base-pairs
effectively behave as rigid entities, the results of El Hassan and
Calladine~\cite{ElHassanCalladine_ptrs_97} and of Hunter et
al.~\cite{HunterLu_jmb_97,Hunter_jmb_92} suggest that the dinucleotide
parameters observed in oligomer crystals can be understood as a
consequence of van-der-Waals and electrostatic interactions between
the neighboring base-pairs and constraints imposed by the
sugar-phosphate backbone.

The purpose of the present paper is the introduction of a class of
``DNA-like''-molecules with simplified interactions resolved at the
base or base pair level. In order to represent the stacking
interactions between neighboring bases (base pairs) we use a
variant~\cite{ralf_GB} of the Gay-Berne potential~\cite{Gay-Berne}
used in studies of discotic liquid crystals. The sugar-phosphate
backbones are reduced to semi-rigid springs connecting the edges of
the disks/ellipsoids.  Using Monte-Carlo simulations we explore the
local stacking and the global helical properties as a function of the
model parameters. In particular, we measure the effective parameters
needed to describe our systems in terms of stack-of-plates (SOP) and
worm-like chain models respectively. This allows us to construct
DNA models which faithfully represent the equilibrium
structure, fluctuations and linear response. At the same time we
preserve the possibility of local structural transitions, e.g. in
response to external forces.

The paper is organized as follows. In the second section we introduce
the base-pair parameters to discuss the helix geometry in terms of
these variables. Furthermore we discuss how to translate the base-pair
parameters in macroscopic variables such as bending and torsional
rigidity. In the third section we introduce the model and discuss the
methods (MC simulation, energy minimization) that we use to explore
its behavior. In the fourth section we present the resulting
equilibrium structures, the persistence lengths as a function of the
model parameters, and the behavior under stretching.

\section{Theoretical Background}

\subsection{Helix geometry}
\label{sec:geo}

\begin{figure}[t]
  \begin{center}
    \includegraphics[angle=0,width=0.90\linewidth]{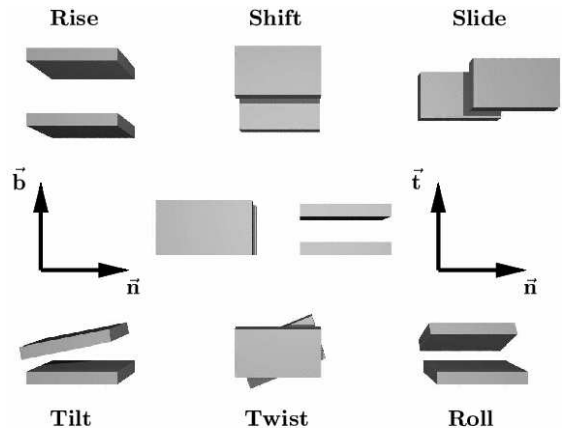}
    \end{center} \caption{(Color online) Illustration of all six base-pair parameters
      and the corresponding coordinate system.} \label{fig:bp}
\end{figure}

To resolve and interpret X-ray diffraction studies on DNA oligomers
the relative position and orientation of successive base-pairs are
analyzed in terms of Rise (Ri), Slide (Sl), Shift (Sh), Twist (Tw),
Roll (Ro), and Tilt (Ti)~\cite{Olson_jmb_94b} (see Fig. \ref{fig:bp}).
In order to illustrate the relation between these local parameters and
the overall shape of the resulting helix we discuss a simple
geometrical model where DNA is viewed as a twisted ladder where all
bars lie in one plane. For vanishing bending angles with
$\mbox{Ro}=\mbox{Ti}=0$
each step is characterized by four parameters: Ri, Sl, Sh, and
Tw~\cite{CalladineDrew99}. Within the given geometry a base pair can
be characterized by its position $\mathbf{r}$ and the angle of its
main axis with the $\mathbf{n}$/$\mathbf{b}$-axis ($\mathbf{n}$ points
into the direction of the large axis, $\mathbf{b}$ points into the
direction of the small axis, and $\mathbf{t}$, representing the
tangent vector of the resulting helix, is perpendicular to the
$\mathbf{n}$-$\mathbf{b}$- plane as it is illustrated in
Fig. \ref{fig:bp}). At each step the center points are displaced by
a distance $\sqrt{\mbox{Sl}^2+\mbox{Sh}^2}$ in the
$\mathbf{n}-\mathbf{b}-$plane. The angle between successive steps is
equal to the twist angle and the center points are located on a helix
with radius $r=\sqrt{\mbox{Sl}^2+\mbox{Sh}^2}/(2\sin(\mbox{Tw}/2))$.

In the following we study the consequences of imposing a simple
constraint on the bond lengths $l_1$ and $l_2$ representing the two
sugar phosphate backbones (the rigid bonds connect the right and left
edges of the bars along the $\mathbf{n}$-axis respectively). Ri is the
typical height of a step which we will try to impose on the grounds
that it represents the preferred stacking distance of neighboring base
pairs. We choose $\mbox{Ri}=3.3\mbox{\AA}$ corresponding to the B-DNA
value. One possibility to fulfill the constraint
$l_1=l_2=l=6\mbox{\AA}$ is pure twist. In this case a relationship of
the twist angle and the width of the base-pairs $d$, the backbone
length $l$ and the imposed rise is obtained:
\begin{equation}
\label{tw:geo}
\mbox{Tw} = \arccos\left(\frac{d^2-2l^2+2\mbox{Ri}^2}{d^2}\right).
\end{equation}
Another possibility is to keep the rotational orientation of the base
pair ($\mbox{Tw}=0$), but to displace its center in the
$\mathbf{n}$-$\mathbf{b}$-plane, in which case
$\mbox{Ri}^2+\mbox{Sl}^2+\mbox{Sh}^2 \equiv l^2$. With $\mbox{Sh}=0$,
it results in a skewed ladder with skew angle
$\arcsin(\mbox{Sl}/l)/\pi$ \cite{CalladineDrew99}.

The general case can be solved as well. In a first step a general
condition is obtained that needs to be fulfilled by any combination of
Sh, Sl, and Tw independently of Ri. For non-vanishing Tw this yields a
relation between Sh and Sl:
\begin{equation}
\label{eq:slsh}
\tan(\mbox{Tw}) = \frac{\mbox{Sh}}{\mbox{Sl}}.
\end{equation}
Using Eq.~(\ref{eq:slsh}) the general equation can finally be solved:
\begin{equation}
\label{twsl}
 \mbox{Sl} = \frac{1}{\sqrt{2}}\left[
    \cos(\frac{\mbox{Tw}}{2})^2
    \sqrt{\sec(\frac{\mbox{Tw}}{2})^2(2l^2-d^2-\mbox{Ri}^2)}\,\right].
\end{equation}
Eq.~(\ref{twsl}) is the result of the mechanical coupling of slide,
shift and twist due to the backbones. Treating the rise again as a
constraint the twist is reduced for increasing slide or shift motion.
\begin{figure}[t]
  \begin{center}
    \includegraphics[angle=0,width=0.99\linewidth]{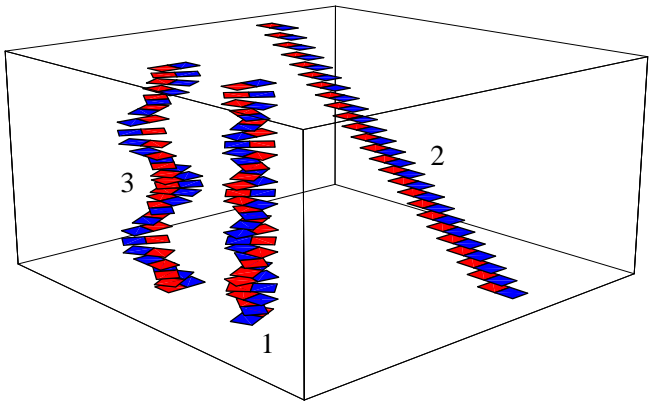}
  \end{center}
\caption{(Color online) Illustration of DNA geometry for a diameter of
  $d=16\mbox{\AA}$: (1) Twisted ladder with $\mbox{Sl}=\mbox{Sh}=0$,
  $\mbox{Ri}=3.3\mbox{\AA}$, $\mbox{Tw}\approx2\pi/10$, (2) Skewed
  ladder with $\mbox{Tw}=\mbox{Sh}=0$, $\mbox{Ri}=3.4\mbox{\AA}$,
  $\mbox{Sl}\approx5.0\mbox{\AA}$, (3) Helix with $\mbox{Tw}=2\pi/12$,
  $\mbox{Ri}=3.4\mbox{\AA}$, $\mbox{Sl}\approx2.7\mbox{\AA}$,
  $\mbox{Sh}\approx1.6\mbox{\AA}$.}
\label{fig:geo}
\end{figure}
The center-center distance $c$ of two neighboring base-pairs is given by
\begin{equation}
  \label{ctwsl}
  c = \sqrt{\mbox{Ri}^2+\mbox{Sl}^2
    \left(
      1+\tan(\mbox{Tw})^2
    \right)}.
\end{equation}
For $\mbox{Tw}=0$ and a given value of $\mbox{Ri}$ the center-center
distance is equal to the backbone length $l$ and for
$\mbox{Tw}=\arccos\left((d^2-2l^2+2\mbox{Ri}^2)/d^2\right)$ one
obtains $c=\mbox{Ri}$.

\subsection{Thermal fluctuations}

In this section we discuss how to calculate the effective coupling
constants of a harmonic system valid within linear response theory
describing the couplings of the base-pair parameters along the
chain. Furthermore we show how to translate measured mean and mean
squared values of the 6 microscopic base-pair parameters into
macroscopic observables such as bending and torsional persistence
length. This provides the linkage between the two descriptions: WLC
(worm-like chain) versus SOP (stack-of-plates) model.

Within linear response theory it should be possible to map our model
onto a Gaussian system where all translational and rotational degrees
of freedom are harmonically coupled. We refer to this model as the
stack-of-plates (SOP) model~\cite{Hern_epjb_98}. The effective
coupling constants are given by the second derivatives of the free
energy in terms of base-pair variables around the equilibrium
configuration. This yields $6\times6$ matrices ${\cal K}^{nm}$
describing the couplings of the base-pair parameters of neighboring
base-pairs along the chain:
\begin{equation}
{\cal K}^{nm} = \frac{\partial^2{\cal F}}{\partial x^n_i\partial x^m_j}.
\end{equation}
Therefore one can calculate the $(N-1)\times(N-1)$ correlation matrix
${\cal C}$ in terms of base-pair parameters. $N$ is thereby the number
of base-pairs.
\begin{equation}
\label{eq:sopcorr}
\langle{\cal C}\rangle =
\begin{pmatrix}
{\cal K}^{11} & {\cal K}^{12} & {\cal K}^{13} & {\cal K}^{14} & \ldots \\
{\cal K}^{12} & {\cal K}^{22} & {\cal K}^{23} & {\cal K}^{24} & \ldots \\
 &  & \ddots &  &
\end{pmatrix}^{-1}.
\end{equation}
The inversion of ${\cal C}$ results in a generalized connectivity
matrix with effective coupling constants as entries.

The following considerations are based on the assumption that one only
deals with nearest-neighbor interactions. Then successive base-pair
steps are independent of each other and the calculation of the
orientational correlation matrix becomes feasible. In the absence of
spontaneous displacements ($\mbox{Sl}=\mbox{Sh}=0$) and spontaneous
bending angles ($\mbox{Ti}=\mbox{Ro}=0$) as it is the case for B-DNA
going from one base-pair to the neighboring implies three operations.
In order to be independent of the reference base pair one first
rotates the respective base pair into the mid-frame with
${\cal{R}}(Tw_{sp}/2)$ (${\cal{R}}$ is a rotation matrix, $Tw_{sp}$
denotes the spontaneous twist), followed by a subsequent overall
rotation in the mid-frame
\begin{equation}
{\cal{A}} =
\begin{pmatrix} {\mathbf{t_i}}\cdot{\mathbf{t_{i+1}}} & {\mathbf{t_i}}\cdot{\mathbf{b_{i+1}}} &
    {\mathbf{t_i}}\cdot{\mathbf{n_{i+1}}} \\
    {\mathbf{b_i}}\cdot{\mathbf{t_{i+1}}} &
    {\mathbf{b_i}}\cdot{\mathbf{b_{i+1}}} &
    {\mathbf{b_i}}\cdot{\mathbf{n_{i+1}}} \\
    {\mathbf{n_i}}\cdot{\mathbf{t_{i+1}}} &
    {\mathbf{n_i}}\cdot{\mathbf{b_{i+1}}} &
    {\mathbf{n_i}}\cdot{\mathbf{n_{i+1}}}
\end{pmatrix}
\end{equation}
taken into account the thermal motion of Ro, Ti and Tw, and a final
rotation due to the spontaneous twist ${\cal{R}}(Tw_{sp}/2)$. The
orientational correlation matrix between two neighboring base pairs
can be written as
$\langle{\cal{O}}_{i\,i+1}\rangle={\cal{R}}(Tw_{sp}/2)\,\langle{\cal{A}}\rangle\,{\cal{R}}(Tw_{sp}/2)$.
${\cal{A}}$ describes the fluctuations around the mean values.  As a
consequence of the independence of successive base-pair parameters one
finds
$\langle{\cal{O}}_{i\,j}\rangle=\left({\cal{R}}(Tw_{sp}/2)\,\langle{\cal{A}}\rangle\,{\cal{R}}(Tw_{sp}/2)\right)^{j-i}$
where the matrix product is carried out in the eigenvector basis of
${\cal{R}}(Tw_{sp}/2)\,\langle{\cal{A}}\rangle\,{\cal{R}}(Tw_{sp}/2)$.
In the end one finds a relationship of the mean and mean squared local
base-pair parameters and the bending and torsional persistence length.
The calculation yields an exponentially decaying tangent-tangent
correlation function
$\langle\mathbf{t}(0)\cdot\mathbf{t}(s)\rangle=\exp(-s/l_p)$ with a
bending persistence length
\begin{equation}
\label{eq:lp}
l_p = \frac{2\langle \mbox{Ri}\rangle}{(\langle \mbox{Ti}^2\rangle + \langle \mbox{Ro}^2\rangle)}.
\end{equation}

In the following we will calculate the torsional persistence length.
Making use of a simple relationship between the local twist and the
base-pair orientations turns out to be more convenient than the
transfer matrix approach.

The (bi)normal-(bi)normal correlation function is an exponentially
decaying function with an oscillating term depending on the helical
repeat length $h=p\langle\mbox{Ri}\rangle$ and the helical pitch
$p=2\pi/\langle\mbox{Tw}\rangle$ respectively, namely
$\langle\mathbf{n}(0)\cdot\mathbf{n}(s)\rangle=\exp(-s/l_n)\cos(2\pi\,s/h)$.
The torsional persistence length $l_n = l_b$ can be calculated in the
following way. It can be shown that the twist angle $\mbox{Tw}$ of two
successive base-pairs is related to the orientations
$\{{\mathbf{t}},{\mathbf{b}},{\mathbf{n}}\}$ and
$\{{\mathbf{t}'},{\mathbf{b}'},{\mathbf{n}'}\}$ through
\begin{equation}
  \cos(\mbox{Tw})=\frac{{\mathbf{n}}\cdot{\mathbf{n}'}+{\mathbf{b}}\cdot{\mathbf{b}'}}{1+{\mathbf{t}}\cdot{\mathbf{t}'}}.
\end{equation}
Taking the mean and using the fact that the orientational correlation
functions and twist correlation function decay exponentially
\begin{equation}
\exp(-1/l_{Tw})=\frac{2\exp(-1/l_n)}{1+\exp(-1/l_p)}
\end{equation}
yields in the case of stiff filaments a simple expression of $l_n$
depending on $l_p$ and $l_{Tw}$:
\begin{equation}
\label{eq:ln}
\frac{l_n}{2} = \frac{l_b}{2} = \left(\frac{2}{l_{Tw}}+\frac{1}{l_p}\right)^{-1},
\end{equation}
where the twist persistence length is defined as
\begin{equation}
\label{eq:ltw}
l_{Tw} = \frac{\langle\mbox{Ri}\rangle}{\langle\mbox{Tw}^2\rangle}.
\end{equation}

\section{Model and methods}

Qualitatively the geometrical considerations suggest a B-DNA like
ground state and the transition to a skewed ladder conformation
under the influence of a sufficiently high stretching force,
because this provides the possibility to lengthen the chain and to
partially conserve stacking. Quantitative modeling requires the
specification of a Hamiltonian.

\subsection{Introduction of the Hamiltonian}

The observed conformation of a dinucleotide base-pair step represents
a compromise between (i) the base stacking interactions (bases are
hydrophobic and the base-pairs can exclude water by closing the gap in
between them) and (ii) the preferred backbone conformation (the
equilibrium backbone length restricts the conformational space
accessible to the base-pairs)~\cite{PackerHunter_jmb_98}. Packer and
Hunter~\cite{PackerHunter_jmb_98} have shown that roll, tilt and rise
are backbone-independent parameters. They depend mainly on the
stacking interaction of successive base-pairs. In contrast twist is
solely controlled by the constraints imposed by a rigid backbone.
Slide and shift are sequence-dependent. While it is possible to
introduce sequence dependant effects into our model, they are ignored
in the present paper.

\begin{figure}[t]
  \begin{center}
    \includegraphics[angle=0,width=0.99\linewidth]{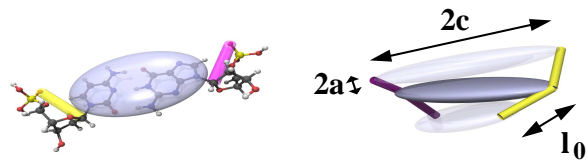}
    \end{center} \caption{(Color online) (left) Illustration of the underlying
    idea. The base-pairs are represented as rigid ellipsoids. The
    sugar-phosphate backbone is treated as semi-rigid springs
    connecting the edges of the ellipsoid. (right) Introduced
    interactions lead to a right-handed twisted structure.}
    \label{fig_model}
\end{figure}

In the present paper we propose a generic model for DNA where the
molecule is described as a stack of thin, rigid ellipsoids
representing the base pairs (Fig. \ref{fig_model}). The shape of the
ellipsoids is given by three radii $a$, $b$, $c$ of the main axes in
the body frames which can be used to define a structure matrix
\begin{equation}
\label{eq:str}
{\cal{S}} =
\begin{pmatrix}
a & 0 & 0 \\
0 & b & 0 \\
0 & 0 & c
\end{pmatrix}
\end{equation}
$2a$ corresponds to the thickness, $2b$ to the depth which is a free
parameter in the model, and $2c=18\mbox{\AA}$ to the width of the
ellipsoid which is fixed to the diameter of a B-DNA helix. The
thickness $2a$ will be chosen in such a way that the minimum
center-center distance for perfect stacking reproduces the
experimentally known value of $3.3\mbox{\AA}$.

The attraction and the excluded volume between the base pairs is
modeled by a variant of the Gay-Berne potential~\cite{ralf_GB,Gay-Berne}
for ellipsoids of arbitrary shape ${\cal S}_i$, relative position
${\vec r}_{1 2}$ and orientation ${\bf A}_i$. The potential can be
written as a product of three terms:
\begin{align}
  \label{eq:GB}
  U({\bf A}_1,{\bf A}_2, {\vec r}_{1 2}) &=
  U_{\mathrm r}({\bf A}_1,{\bf A}_2, {\vec    r}_{1 2}) \nonumber \\
 &\times \eta_{1 2} ({\bf A}_1,   {\bf A}_2,{\hat{r}}_{1 2})
    \,    \chi_{1 2} ({\bf A}_1,{\bf A}_2,{\hat{r}}_{1 2}).
\end{align}
The first term controls the distance dependence of the interaction and
has the form of a simple LJ potential
\begin{equation}\label{eq:U0}
U_{\mathrm r} = 4\epsilon_{\mathrm GB}
\left( \left(\frac\sigma {h+\gamma\sigma}\right)^{12} -
       \left(\frac\sigma {h+\gamma\sigma}\right)^{6}
\strut\right)
\end{equation}
where the interparticle distance $r$ is replaced by the distance
$h$ of closest approach between the two bodies:
\begin{equation}\label{eq:h}
h \equiv \min(|\vec r_i-\vec r_j|) \ \forall (i,j)
\end{equation}
with $i\in{\mathrm Body\, 1}$ and $j\in{\mathrm Body\, 2}$. The range
of interaction is controlled by an atomistic length scale
$\sigma=3.3\mbox{\AA}$, representing the effective diameter of a
base-pair.

In general, the calculation of $h$ is non-trivial. We use the
following approximative calculation scheme which is usually employed
in connection with the Gay-Berne potential:
\begin{eqnarray}
h({\bf A}_1,{\bf A}_2, {\vec r}_{1 2}) &=&
  r_{1 2} -\sigma_{1 2}({\bf A}_1,{\bf A}_2, \hat{r}_{1 2})
\label{eq:h_GB}\\
\sigma_{1 2}({\bf A}_1,{\bf A}_2, \hat{r}_{1 2}) &=& [{\frac{1}{2}}
 \hat{r}_{1 2}^T {\bf G}_{1 2}^{-1}({\bf A}_1, {\bf A}_2)
 \hat{r}_{1 2}]^{-1/2}
\label{eq:sigma_12}\\
{\bf G}_{1 2}({\bf A}_1, {\bf A}_2) &=&{\bf A}_1^T {\bf S}_1^2
{\bf A}_1 + {\bf A}_2^T {\bf S}_2^2  {\bf A}_2. \label{eq:G_GB}
\end{eqnarray}
In the present case of oblate objects with rather perfect stacking
behavior Eq.~(\ref{eq:h_GB}) produces only small deviations from the
exact solution of Eq.~(\ref{eq:h}).

The other two terms in Eq.~(\ref{eq:GB}) control the interaction
strength as a function of the relative orientation
${\bf A}_1^t {\bf A}_2$ and position $\vec{r}_{12}$ of interacting
ellipsoids:
\begin{eqnarray}
 \eta_{1 2} ({\bf A}_1, {\bf A}_2,{\hat{r}}_{1 2}) &=&
   \frac{ \det[{\bf S}_1]/\sigma_1^{2}+
          \det[{\bf S}_2]/\sigma_2^{2}
        }
        {\left(\det[{\bf H}_{12}]/(\sigma_1+\sigma_2)\right)^{1/2}} \\
{\bf H}_{1 2}({\bf A}_1,{\bf A}_2, \hat{r}_{1 2}) &=&
\frac1{\sigma_1}{\bf A}_1^T {\bf S}_1^2  {\bf A}_1 +
\frac1{\sigma_2}{\bf A}_2^T {\bf S}_2^2  {\bf A}_2\\
\sigma_i({\bf A}_i, \hat{r}_{1 2}) &\equiv&
\left(\hat{r}_{1 2}^T\ \ {\bf A}_1^T {\bf S}_{i}^{-2}{\bf A}_1 \ \
 \hat{r}_{1 2}\right)^{-1/2}
\end{eqnarray}
and
\begin{eqnarray}
\label{eq:chi_GB}
 \chi_{1 2}({\bf A}_1,{\bf A}_2,  \hat{r}_{1 2}) &=& [2 \hat{r}_{1 2}^T\ \  {\bf B}_{1 2}^{-1}({\bf A}_1, {\bf A}_2)  \ \hat{r}_{1 2}]\\
\label{eq:B}
  {\bf B}_{1 2}({\bf A}_1, {\bf A}_2) &=&  {\bf A}_1^T {\bf E}_1  {\bf
    A}_1 + {\bf A}_2^T {\bf E}_2  {\bf A}_2
\end{eqnarray}
with
\begin{equation}
{\bf E}_i =  \sigma
\begin{pmatrix} \frac{a_i}{b_i\,c_i} & 0 & 0 \\
                0 & \frac{b_i}{a_i\,c_i} & 0 \\
                0 & 0 & \frac{c_i}{a_i\,b_i}
\end{pmatrix}
= \frac\sigma{\det[{\bf S}_i]} {\bf S}_i^2.
\end{equation}

We neglect electrostatic interactions between neighboring base-pairs
since at physiological conditions the stacking interaction
dominates~\cite{Hunter_jmb_92,CalladineDrew99}.

At this point we have to find appropriate values for the thickness
$2a$ and the parameter $\gamma$ of Eq.~(\ref{eq:U0}). Both
parameters influence the minimum of the Gay-Berne potential. There
are essentially two possible procedures. One way is to make use of
the parameterization result of Everaers and Ejtehadi~\cite{ralf_GB},
i.e. $\gamma=2^{1/6}-30^{-1/6}$, and to choose a value of
$a\approx0.7$ that yields the minimum center-center distance of
$3.3\mbox{\AA}$ for perfect stacking. Unfortunately it turns out that
the fluctuations of the bending angles strongly depend on the
flatness of the ellipsoids. The more flat the ellipsoids are the
smaller are the fluctuations of the bending angles so that one
ends up with extremely stiff filaments with a persistence length
of a few thousand base-pairs. This can be seen clearly for the
extreme case of two perfectly stacked plates: each bending move
leads then to an immediate overlap of the plates. That is why we
choose the second possibility. We keep $\gamma$ as a free
parameter that is used in the end to shift the potential minimum
to the desired value and fix the width of the ellipsoids to be
approximately half the known rise value $a=1.55\mbox{\AA}$. This
requires $\gamma=1.07$.

The sugar phosphate backbone is known to be nearly inextensible. The
distance between adjacent sugars varies from $5.5\mbox{\AA}$ to
$6.5\mbox{\AA}$~\cite{CalladineDrew99}. This is taken into account by
two stiff springs with length $l_1=l_2=6.0\mbox{\AA}$ connecting
neighboring ellipsoids (see Fig. \ref{fig_model}). The anchor points
are situated along the centerline in $\vec{n}$-direction (compare
Fig. \ref{fig:bp} and Fig. \ref{fig_model}) with a distance of
$\pm8\mbox{\AA}$ from the center of mass. The backbone is thus
represented by an elastic spring with non-zero spring length
$l_0=6\mbox{\AA}$
\begin{equation}
{\cal H}_{el} = \frac{k}{2}\left[(|\mathbf{r}_{1,i+1}-\mathbf{r}_{1,i}|-l_0)^2 +
  (|\mathbf{r}_{2,i+1}-\mathbf{r}_{2,i}|-l_0)^2\right].
\end{equation}

Certainly a situation where the backbones are brought closer to one
side of the ellipsoid so as to create a minor and major groove would
be a better description of the B-DNA structure. But it turns out that
due to the ellipsoidal shape of the base-pairs and due to the fact
that the internal base-pair degrees of freedom (propeller twist, etc.)
cannot relax a non-B-DNA-like ground state is obtained where roll and
slide motion is involved.

The competition between the GB potential that forces the ellipsoids to
maximize the contact area and the harmonic springs with non-zero
spring length that does not like to be compressed leads to a twist in
either direction of the order of $\pm\pi/5$. The right-handedness of
the DNA helix is due to excluded volume interactions between the bases
and the backbone~\cite{CalladineDrew99} which we do not represent
explicitly. Rather we break the symmetry by rejecting moves which lead
to local twist smaller than $-\pi/18$.

Thus we are left with three free parameters in our model, the GB
energy depth $\epsilon=\min(U)$ which controls the stacking
interaction, the spring constant $k$ which controls the torsional
rigidity, and the depth $b$ of the ellipsoids which influences mainly
the fluctuations of the bending angles. All other parameters such as
the width and the height of the ellipsoids, or the range of
interaction $\sigma=3.3\mbox{\AA}$ which determines the width of the
GB potential are fixed so as to reproduce the experimental values for
B-DNA.

\subsection{MC simulation}

In our model all interactions are local and it can therefore
conveniently be studied using a MC scheme. In addition to trial moves
consisting of local displacements and rotations of one ellipsoid by a
small amplitude, it is possible to employ global moves which modify
the position and the orientation of large parts of the chain. The
moves are analogous of (i) the well-known pivot move~\cite{Binder_00},
and (ii) a crankshaft move where two randomly chosen points along the
chain define the axis of rotation around which the inner part of the
chain is rotated. The moves are accepted or rejected according to the
Metropolis scheme~\cite{Metropolis_jcp_53}.

\begin{figure}[t]
  \begin{center}
    \includegraphics[angle=0,width=0.99\linewidth]{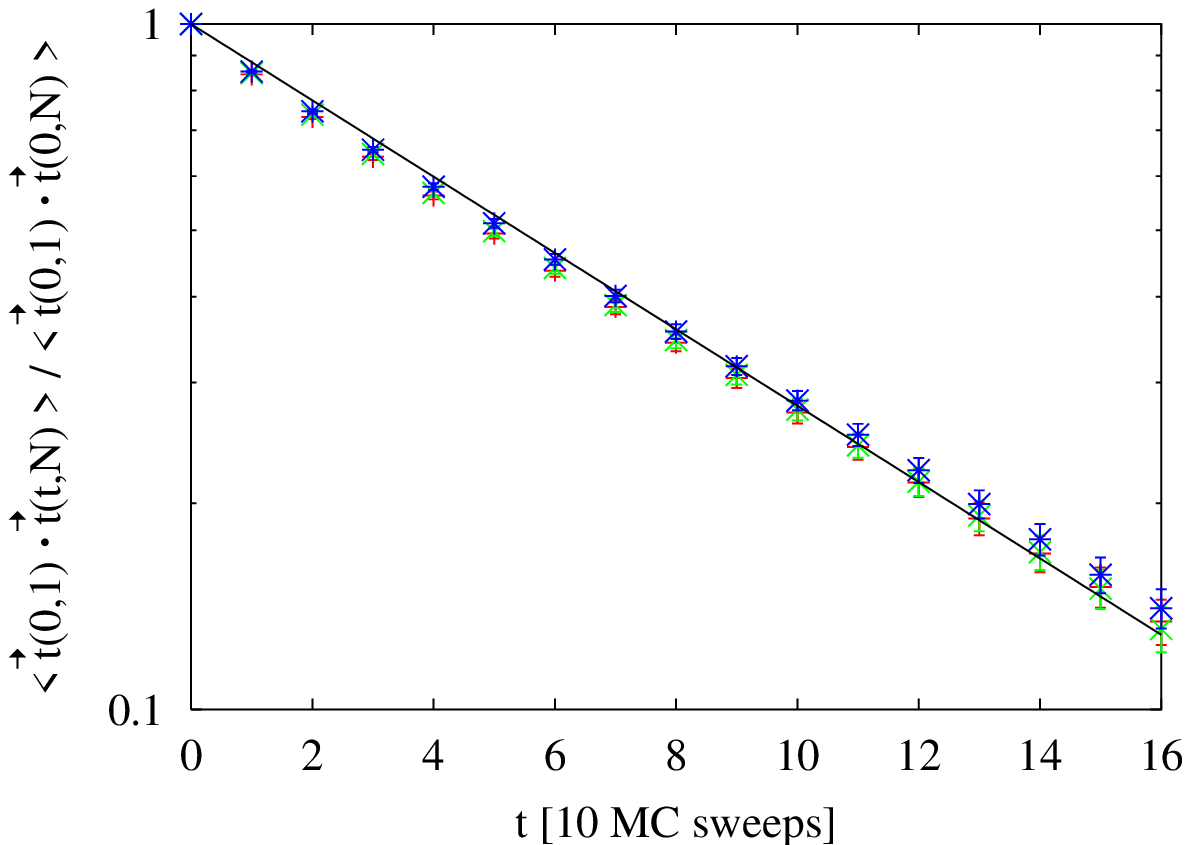}
    \includegraphics[angle=0,width=0.99\linewidth]{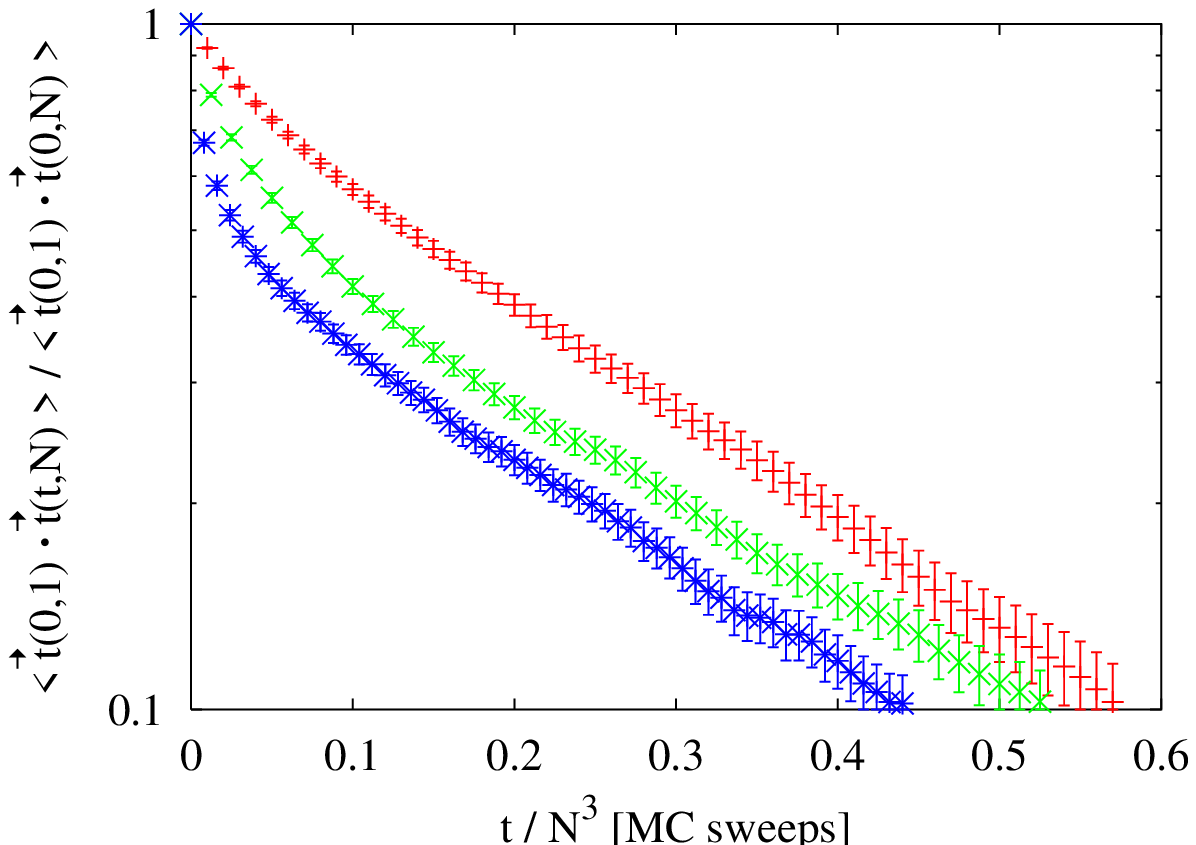}
    \end{center} \caption{(Color online) Time correlation functions of the scalar
      product of the tangent vectors of the first and the last monomer
      $\tau=\vec{t}(0,1)\cdot\vec{t}(t,N)$ with $N=10\mbox{
        (red)},\,N=20\mbox{ (green)},\,N=50\mbox{ (blue)}$ for (a)
      global and (b) local moves. It is observed that $\tau_{global}$
      is independent of the chain length $N$ whereas $\tau_{local}$
      scales as $N^3$.  The 'time' is measured in units of sweeps
      where one MC sweep corresponds to $N$ trials. The CPU time for
      one sweep scales as $N^2$ in case of global moves and as $N$ in
      case of local moves. Thus the simulation time $t$ scales as
      $t_{local}\propto N^4$ and $t_{global}\propto N^2$.}
    \label{fig:moves}
\end{figure}

Fig. \ref{fig:moves} shows that these global moves significantly
improve the efficiency of the simulation. We measured the correlation
time $\tau$ of the scalar product of the tangent vectors of the first
and the last monomer of 200 independent simulation runs with
$N=10,\,20,\,50$ monomers using (i) only local moves and (ii) local
and global moves (ratio 1:1). The correlation time of the global
moves is independent of the chain length with
$\tau_{global}\approx78\,\mbox{ sweeps}$ whereas $\tau_{local}$
scales as $N^3$.

Each simulation run comprises $10^6$ MC sweeps where one MC sweep
corresponds to $2N$ trials (one rotational and one translational move
per base pair) with $N$ denoting the number of monomers.  The
amplitude is chosen such that the acceptance rate equals approximately
to $50${\%}. Every 1000 sweeps we store a snapshot of the DNA
conformation. We measured the 'time' correlation functions of the
end-to-end distance, the rise of one base-pair inside the chain and
all three orientational angles of the first and the last monomer and
of two neighboring monomers inside the chain in order to extract the
longest relaxation time $\tau_{max}$. We observe $\tau_{max}<1000$ for
all simulation runs.

An estimate for the CPU time required for one sweep for chains of
length $N=100$ on a AMD Athlon MP 2000+ processor results in $0.026s$
which is equivalent to $1.33\,10^{-4}s$ per move.

\subsection{Energy minimization}

We complemented the simulation study by zero temperature
considerations that help to discuss the geometric structure that is
obtained by the introduced interactions and to rationalize the MC
simulation data. Furthermore they can be used to obtain an estimate of
the critical force $f_{crit}$ that must be applied to enable the
structural transition from B-DNA to the overstretched S-DNA
configuration as a function of the model parameters
$\{\epsilon,k,b\}$.

\section{Results}

In the following we will try to motivate an appropriate parameter set
$\{\epsilon,\,k,\,b\}$ that can be used for further investigations
within the framework of the presented model. Therefore we explore the
parameter dependence of experimental observables such as the bending
persistence length of B-DNA $l_p\approx150\mbox{bp}$, the torsional
persistence length $l_t\approx260\mbox{bp}$~\cite{Strick_genetica_99},
the mean values and correlations of all six base-pair parameters and
the critical pulling force $f_{crit}\approx65\mbox{pN}$
\cite{Cluzel_sci_96,Lavery_genetica_99,Lavery_jpcm_02,Smith_cosb_00}
that must be applied to enable the structural transition from B-DNA to
the overstretched S-DNA configuration. In fact, static and dynamic
contributions to the bending persistence length $l_p$ of DNA are still
under discussion. It is known that $l_p$ depends on both the intrinsic
curvature of the double helix due to spontaneous bending of particular
base-pair sequences and the thermal fluctuations of the bending
angles. Bensimon {\em et al.}~\cite{Bensimon_epl_98} introduced
disorder into the WLC model by an additional set of preferred random
orientation between successive segments and found the following
relationship between the pure persistence length $l_{pure}$, i.e.
without disorder, the effective persistence length $l_{eff}$ and the
persistence length $l_{disorder}$ caused by disorder:
\begin{equation}
  \frac{l_{eff}}{l_{pure}} = 
  \begin{cases}
    1-\frac{\sqrt{\frac{l_{pure}}{l_{disorder}}}}{2} & \frac{l_{pure}}{l_{disorder}} \ll 1 \\
    \frac{2}{\frac{l_{pure}}{l_{disorder}}} & \frac{l_{pure}}{l_{disorder}} \gg 1
  \end{cases}.
\end{equation}
Since we are dealing with intrinsically straight filaments with
$1/l_{disorder}=0$, we measure $l_{pure}$. Recent estimates of
$l_{disorder}$ range between $430$~\cite{Bednar_jmb_95} and
$4800$~\cite{Vologodskaia_jmb_02} base-pairs using cryo-electron
microscopy and cyclization experiments respectively implicating values
between $105$ and $140$ base-pairs for $l_{pure}$.

\subsection{Equilibrium structure}

As a first step we study the equilibrium structure of our chains as a
function of the model parameters. To investigate the ground state
conformation we rationalize the MC simulation results with the help of
the geometrical considerations and minimum energy calculations. In the
end we will choose parameters for which our model reproduces the
experimental values of B-DNA ~\cite{CalladineDrew99}:
\begin{eqnarray}
\langle\mbox{Ri}\rangle &=& 3.3-3.4\mbox{\AA}\nonumber\\
\langle\mbox{Sl}\rangle &=& 0\mbox{\AA}\nonumber\\
\langle\mbox{Sh}\rangle &=& 0\mbox{\AA}\nonumber\\
\langle\mbox{Tw}\rangle &=& 2\pi/10.5-2\pi/10\nonumber\\
\langle\mbox{Ti}\rangle &=& 0\nonumber\\
\langle\mbox{Ro}\rangle &=& 0.\nonumber
\end{eqnarray}
We use the following reduced units in our calculations. The energy
is measured in units of $k_BT$, lengths in units of $\mbox{{\AA}}$,
forces in units of $k_BT\mbox{{\AA}}^{-1}\approx 40\mbox{pN}$.

\begin{table}[t]
  \center{
    \begin{tabular}{c | c c c c c c c c}
      $\,T\,$ & $\langle\mbox{Ri}\rangle$ & $\langle\mbox{Sh}\rangle$
      & $\langle\mbox{Sl}\rangle$ & $\langle\mbox{Tw}\rangle$ &
      $\langle\mbox{Ti}\rangle$ & $\langle\mbox{Ro}\rangle$ & $\langle
      c\rangle$ & $l_p$  \\ \hline
      0  & \,3.26\, & \,0.0\, & \,0.0\, & \,0.64\, & \,0.0\, & \,0.0\,
      & \,3.26\,  & \, $\infty$ \, \\
      1  & \,3.37\, & \,0.01\, & \,-0.01\, & \,0.62\, & \,0.0\, & \,0.0\, & \,3.47\,  & \,172.8\, \\
      2  & \,3.76\, & \,-0.01\, & \,-0.03\, & \,0.47\, & \,0.0\, & \,0.0\, & \,4.41\,  & \,25.3\, \\
      3  & \,4.10\, & \,-0.01\, & \,0.01\, & \,0.34\, & \,0.0\, & \,-0.01\, & \,5.07\,   & \,14.4\, \\
      5  & \,4.30\, & \,0.03\, & \,-0.02\, & \,0.27\, & \,0.0\, & \,0.01\,  & \,5.39\,  & \,13.6\, \\
    \end{tabular}
    \caption{Dependence of mean values of all six step parameters and
      of the mean center-center distance $\langle c\rangle$ on the
      temperature for $2b=11\mbox{\AA}$, $\epsilon=20k_BT$,
      $k=64k_BT/\mbox{\AA}^2$. $\langle\mbox{Ri}\rangle$,
      $\langle\mbox{Sh}\rangle$, $\langle\mbox{Sl}\rangle$ and $\langle
      c\rangle$ are measured in [$\AA$], $l_p$ in base-pairs.}}
  \label{tab:kT}
\end{table}

We start by minimizing the energy for the various conformations shown
in Fig. \ref{fig:geo} to verify that our model Hamiltonian indeed
prefers the B-Form. Since we have only local (nearest neighbor)
interactions we can restrict the calculations to two base pairs. There
are three local minima which have to be considered: (i) a stacked,
twisted conformation with
$\mbox{Ri}=3.3,\,\mbox{Sl,\,Sh,\,Ti,\,Ro}=0,\,\mbox{Tw}=\pi/10$, (ii)
a skewed ladder with $\mbox{Ri}=3.3,\,\mbox{Sl}=
5.0,\,\mbox{Sh,\,Tw,\,Ti,\,Ro}=0$, and (iii) an unwound helix with
$\mbox{Ri}=6.0,\,\mbox{Sl,\,Sh,\,Ti,\,Ro}=0,\,\mbox{Tw}=0$. Without an
external pulling force the global minimum is found to be the stacked
twisted conformation.

We investigated the dependence of Ri and Tw on the GB energy depth
$\epsilon$ that controls the stacking energy for different spring
constants $k$. Ri depends neither on $\epsilon$ nor on $k$ nor on
$b$. It shows a constant value of $\mbox{Ri}\approx3.3\mbox{\AA}$ for all
parameter sets $\{\epsilon,k,b\}$. The resulting Tw of the minimum
energy calculation coincides with the geometrically determined value
under the assumption of fixed Ri up to a critical $\epsilon$. Up to
that value the springs behave effectively as rigid rods. The critical
$\epsilon$ is determined by the torque $\tau(k,\epsilon)$ that has to
be applied to open the twisted structure for a given value of Ri.

\begin{figure}[t]
  \begin{center}
    \includegraphics[angle=0,width=0.99\linewidth]{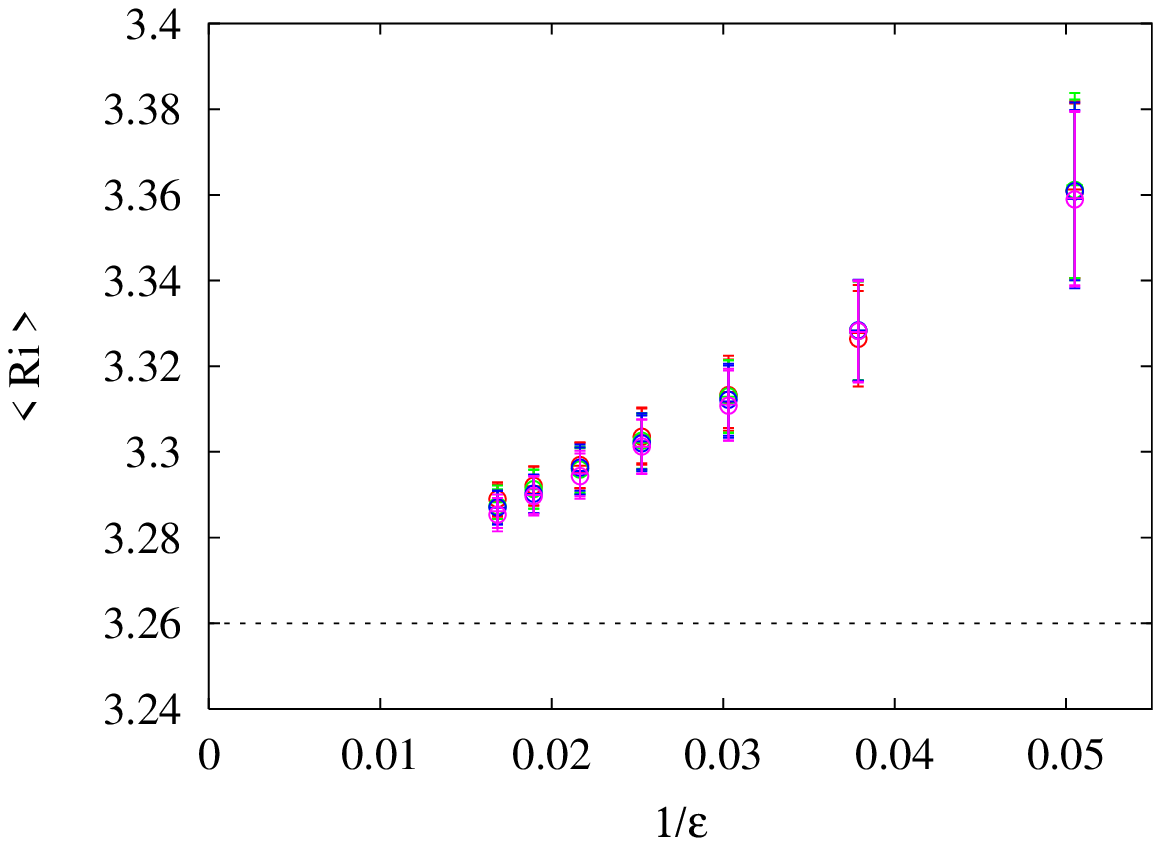}
    \includegraphics[angle=0,width=0.99\linewidth]{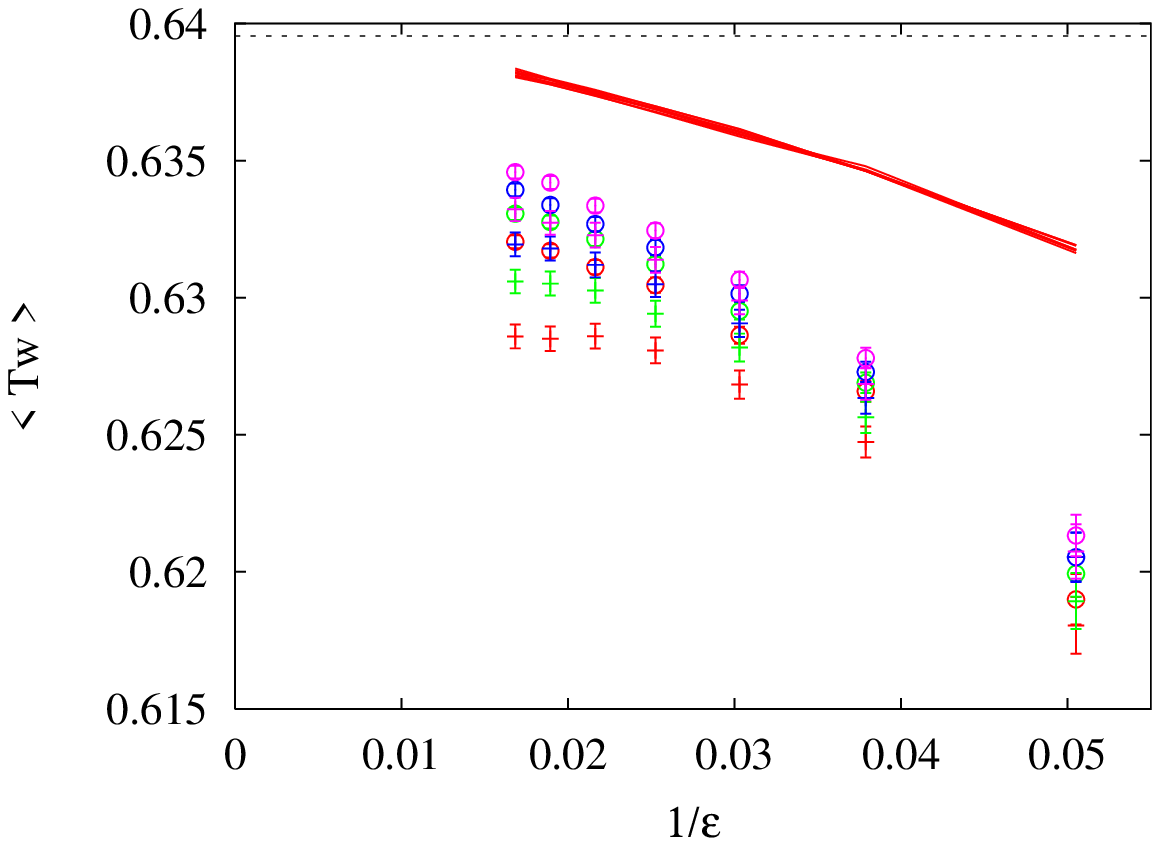}
    \end{center} \caption{(Color online) (a) Rise [$\mbox{\AA}$] and (b) twist 
      as a function of $\epsilon$ [$k_BT$] for
      $2b=8,\,9,\,10,\,11\mbox{\AA}$ (red, green, blue, purple). For
      every $b$ there are two data sets for $k=32\mbox{
        (plus)},\,64\mbox{ (circles)}$ [$k_BT/\mbox{\AA}^2$].
      The dotted line corresponds to the minimum energy value.
      $\langle\mbox{Ri}\rangle$ depends only on $\epsilon$. In the
      limit of $\epsilon\rightarrow\infty$ the minimum energy value is
      reached. \newline (b) In addition to the MC data and the minimum
      energy calculation we calculated the twist with
      Eq.~(\protect{\ref{tw:geo}}) using the measured mean rise values
      of (a). One can observe that $\langle\mbox{Tw}\rangle$ changes
      with all three model parameters. Increasing $y$ and $k$
      decreases especially the fluctuations of $\mbox{Tw}$ and
      $\mbox{Sh}$ so that $\langle\mbox{Tw}\rangle$ increases as a
      result of the mechanical coupling of the shift and twist motion.
      In the limit of $\epsilon,k\rightarrow\infty$ the minimum energy
      value is reached.}\label{fig:tw}
\end{figure}

Using MC simulations we can study the effects arising from thermal
fluctuations. Plotting $\langle\mbox{Ri}\rangle$, and
$\langle\mbox{Tw}\rangle$ as a function of the GB energy depth
$\epsilon$ one recognizes that in general $\langle \mbox{Ri}\rangle$
is larger than $\mbox{Ri}(T=0)$. It converges only for large values of
$\epsilon$ to the minimum energy values. This can be understood as
follows.  Without fluctuations the two base pairs are perfectly
stacked taking the minimum energy configuration
$\mbox{Ri}=3.3\mbox{\AA}$, $\mbox{Sl,\,Sh,\,Ti,\,Ro}=0$, and
$\mbox{Tw}=\pi/10$. As the temperature is increased the fluctuations
can only occur to larger Ri values due to the repulsion of neighboring
base pairs. A decrease of Ri would cause the base-pairs to intersect.
Increasing the stacking energy reduces the fluctuations in the
direction of the tangent vector and leads to smaller $\langle
\mbox{Ri}\rangle$ value. In the limit $\epsilon\rightarrow\infty$ it
should reach the minimum energy value which is observed from the
simulation data. In turn the increase of the mean value of rise
results in a smaller twist angle $\langle \mbox{Tw}\rangle$. We can
calculate with the help of Eq.~(\ref{tw:geo}) the expected twist using
the measured mean values of $\langle\mbox{Ri}\rangle$.  Fig.
\ref{fig:tw} shows that there is no agreement.
\begin{figure}[t]
  \begin{center}
    \includegraphics[angle=0,width=0.8\linewidth]{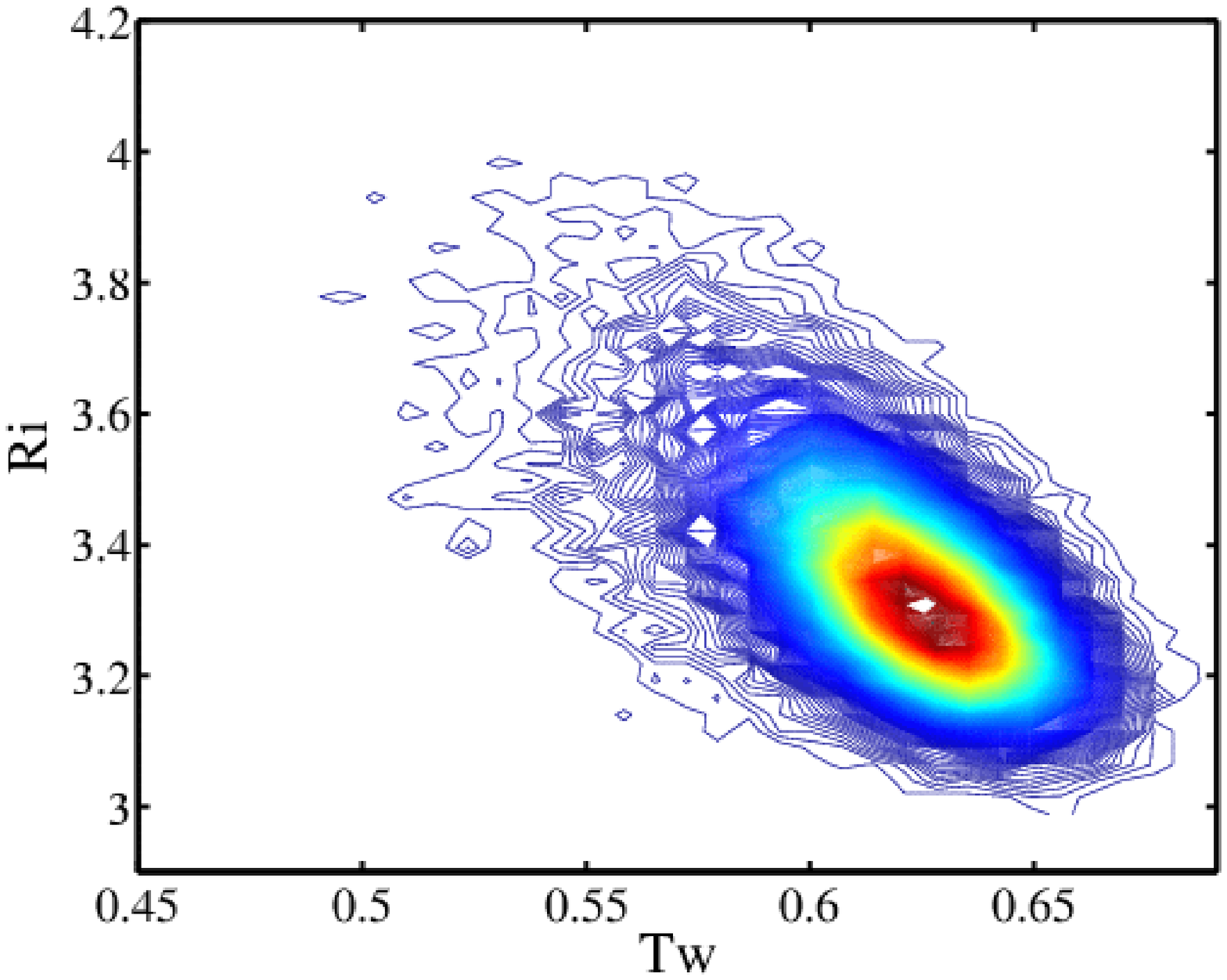}
    \includegraphics[angle=0,width=0.8\linewidth]{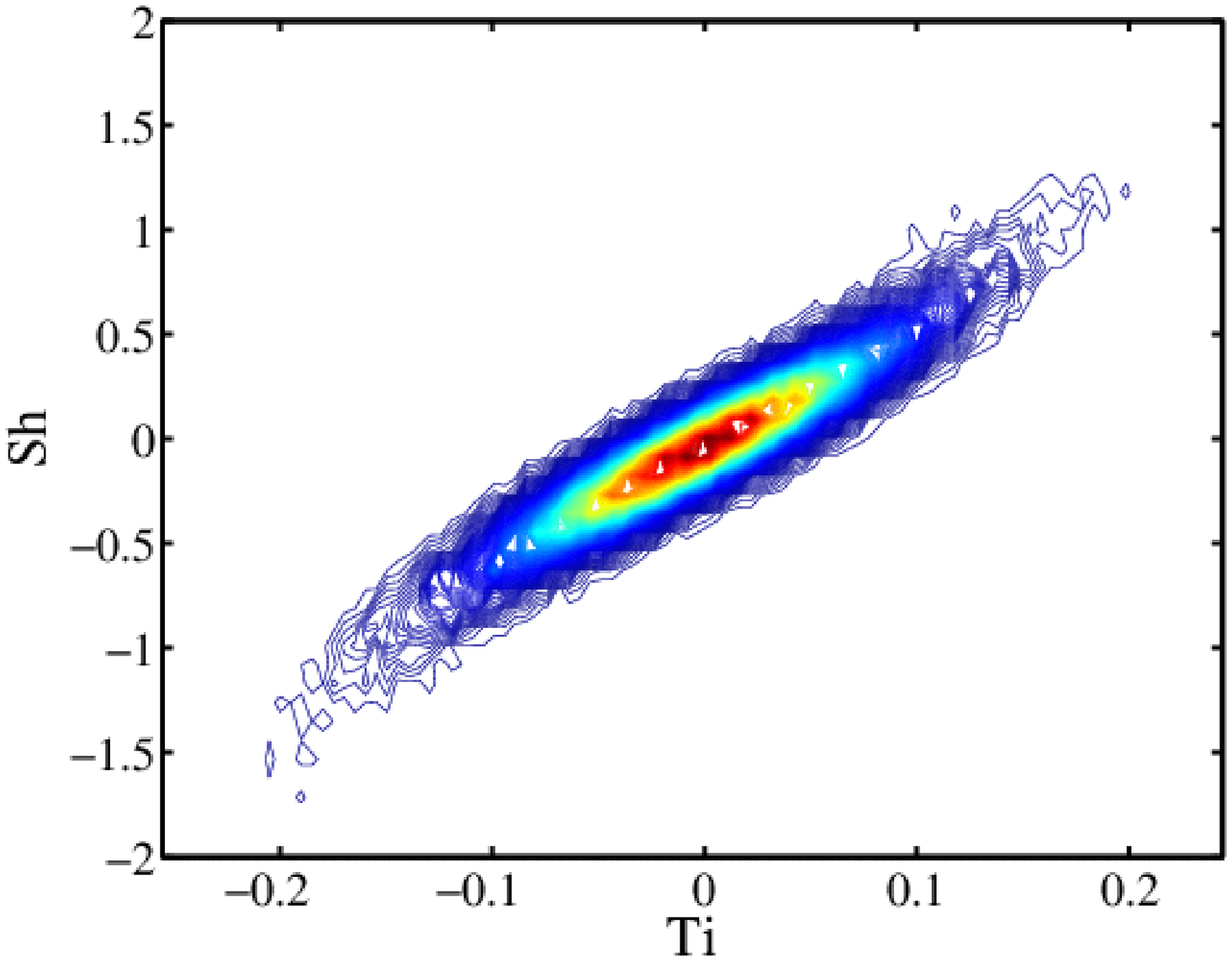}
    \includegraphics[angle=0,width=0.8\linewidth]{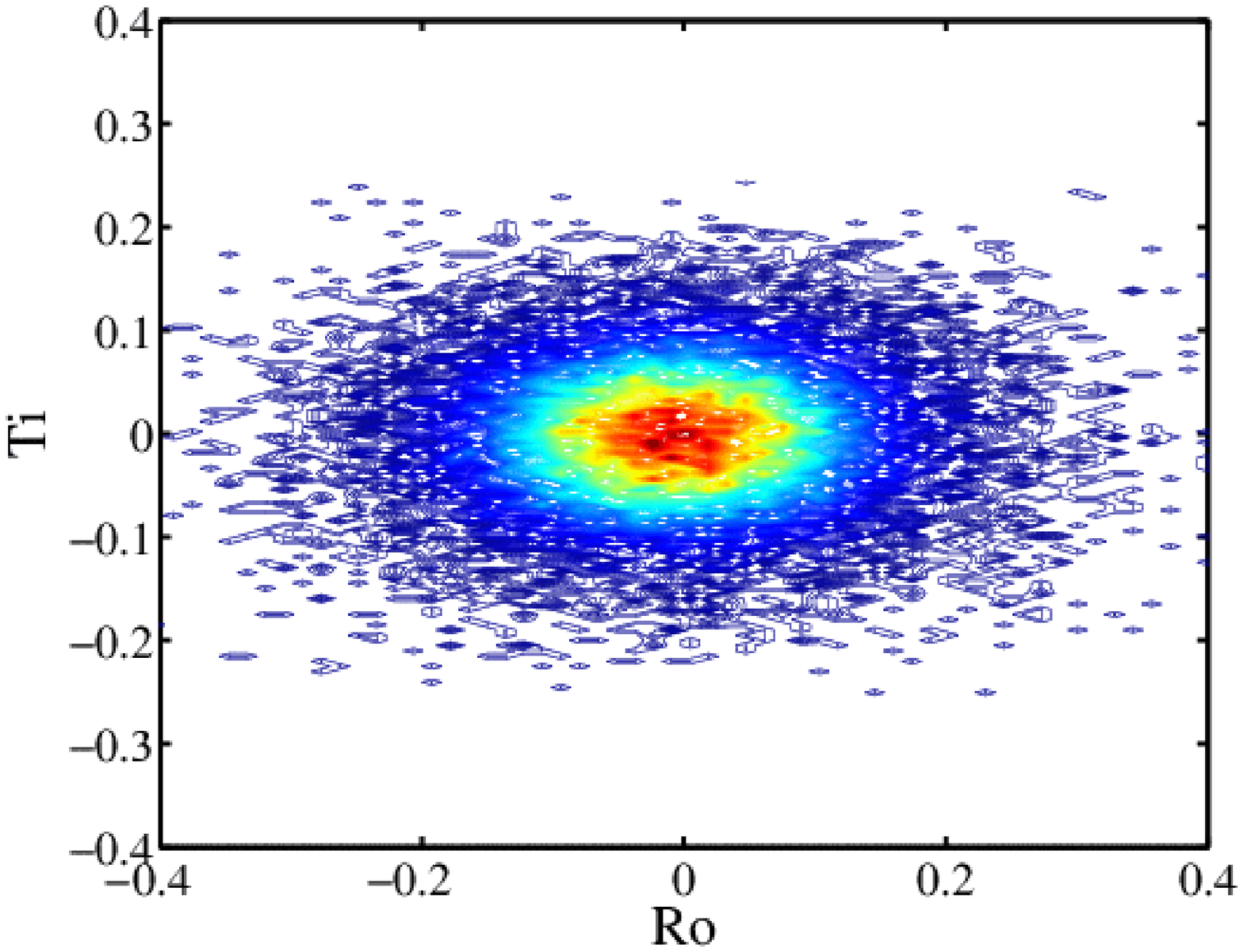}
    \end{center} \caption{(Color online) Contour plots of measured clouds for
      rise-twist, shift-tilt, and roll-tilt to demonstrate internal
      couplings and the anisotropy of the bending angles
      ($2b=11\mbox{\AA}$, $\epsilon=20k_BT$,
      $k=64k_BT/\mbox{\AA}^2$).}
\label{fig:scat}
\end{figure}
The deviations are due to fluctuations in Sl and Sh which cause the
base-pairs to untwist. This is the mechanical coupling of Sl, Sh, and
Tw due to the backbones already mentioned in section \ref{sec:geo}. It
is observed that a stiffer spring $k$ and a larger depth of the
ellipsoids $b$ result in larger mean twist values. Increasing the
spring constant $k$ means decreasing the fluctuations of the twist
and, due to the mechanical coupling, of the shift motion around the
mean values which explains the larger mean twist values. An increase
of the ellipsoidal depth $b$ in turn decreases the fluctuations of the
bending angles.  The coupling of the tilt fluctuations with the shift
fluctuations leads to larger values for $\langle\mbox{Tw}\rangle$.
The corresponding limit where
$\langle\mbox{Tw}\rangle\rightarrow\mbox{Tw}(\mbox{T}=0)$ is given by
$k,\epsilon\rightarrow\infty$.

The measurement of the mean values of all six base-pair step
parameters for different temperatures is shown in Table \ref{tab:kT}.
One can see that with increasing temperature the twist angles decrease
while the mean value of rise increase. The increase of the
center-center distance is not only due to fluctuations in $\mbox{Ri}$
but also due to fluctuations in $\mbox{Sl}$ and $\mbox{Sh}$. That is
why there are strong deviations of $\langle c\rangle$ from
$\langle\mbox{Ri}\rangle$ even though the mean values of $\mbox{Sl}$
and $\mbox{Sh}$ vanish. Note that the mean backbone length $\langle
l\rangle$ always amounts to about $6\mbox{\AA}$.

\begin{figure}[t]
  \begin{center}
    \includegraphics[angle=0,width=0.49\linewidth]{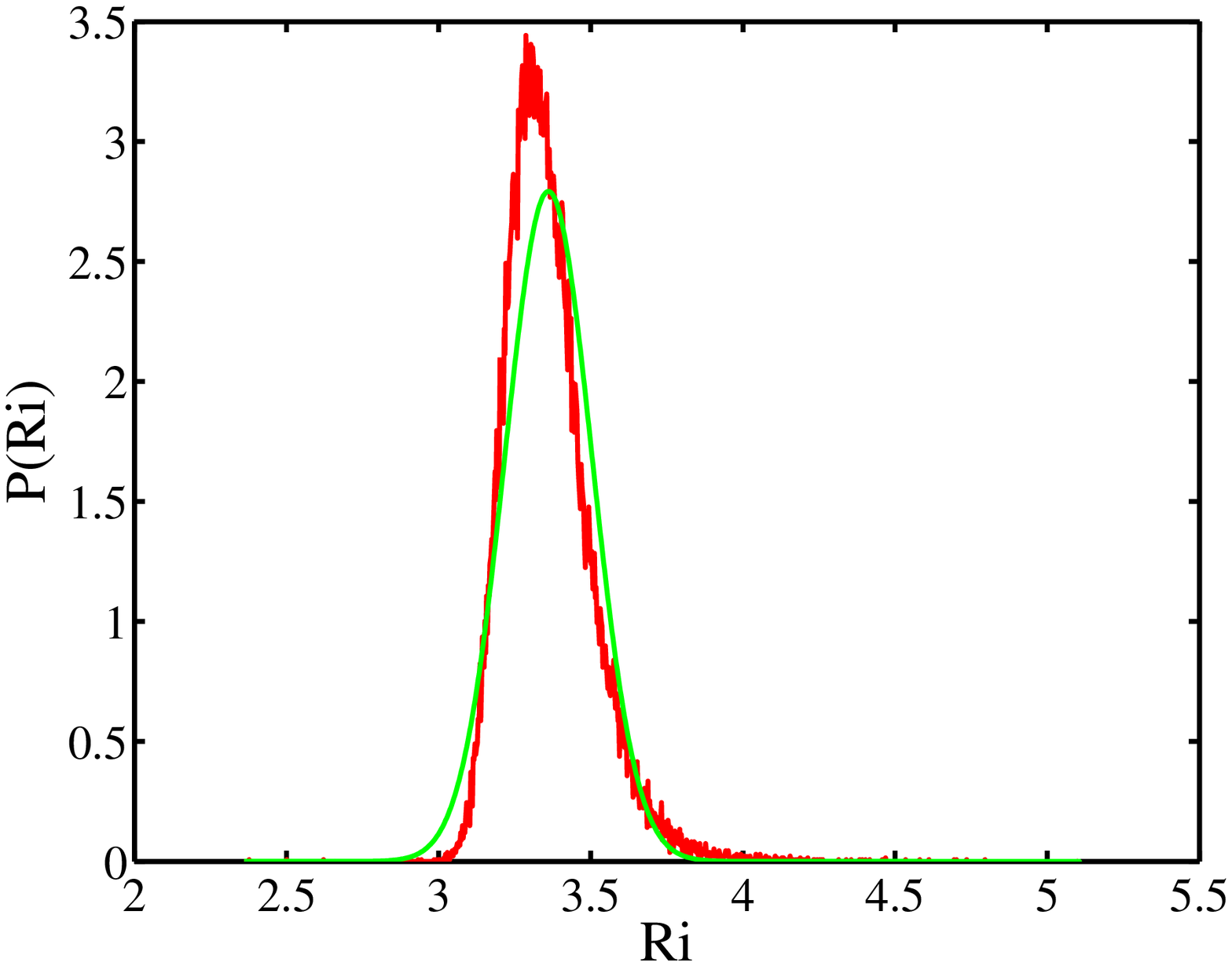}
    \includegraphics[angle=0,width=0.49\linewidth]{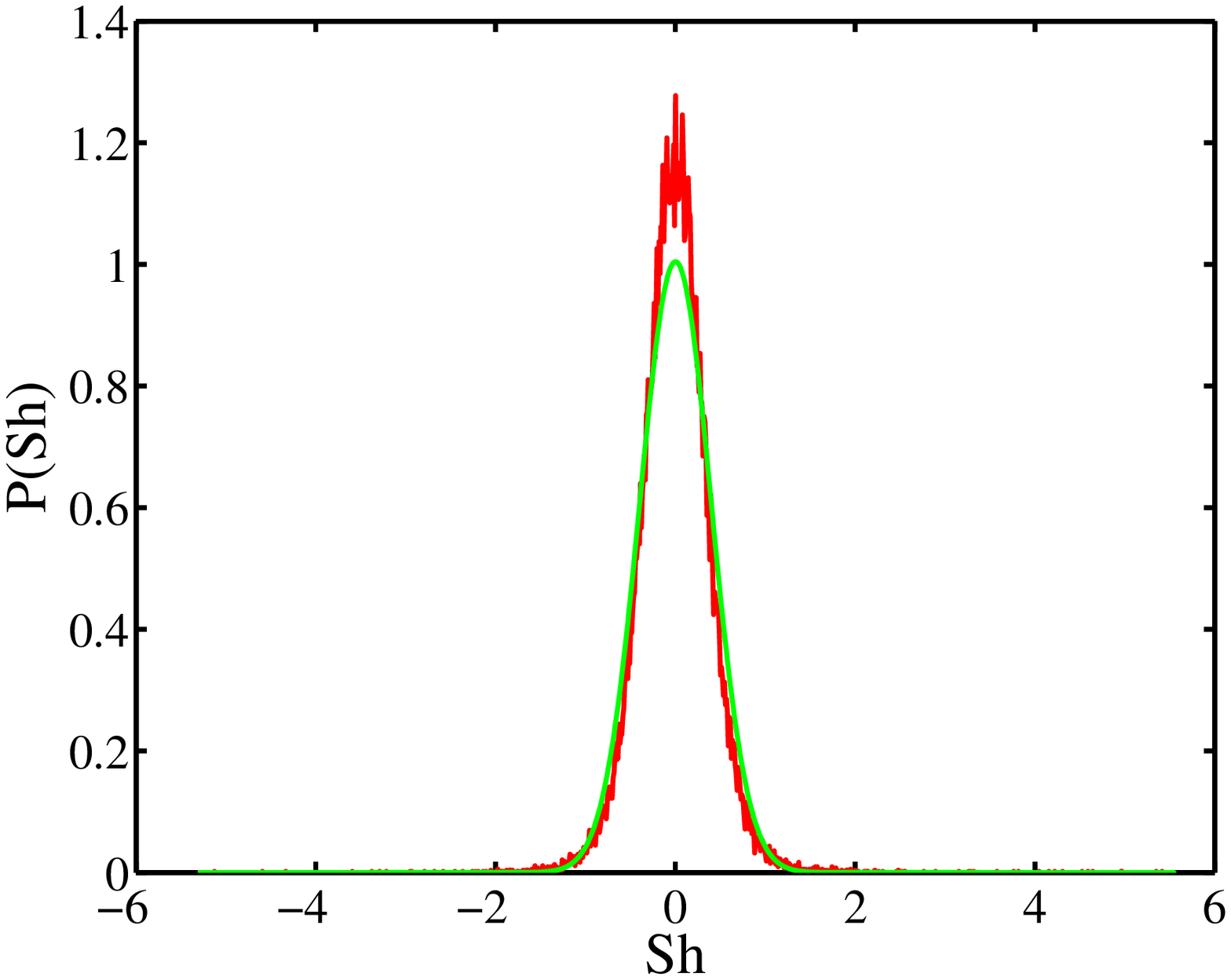}
    \includegraphics[angle=0,width=0.49\linewidth]{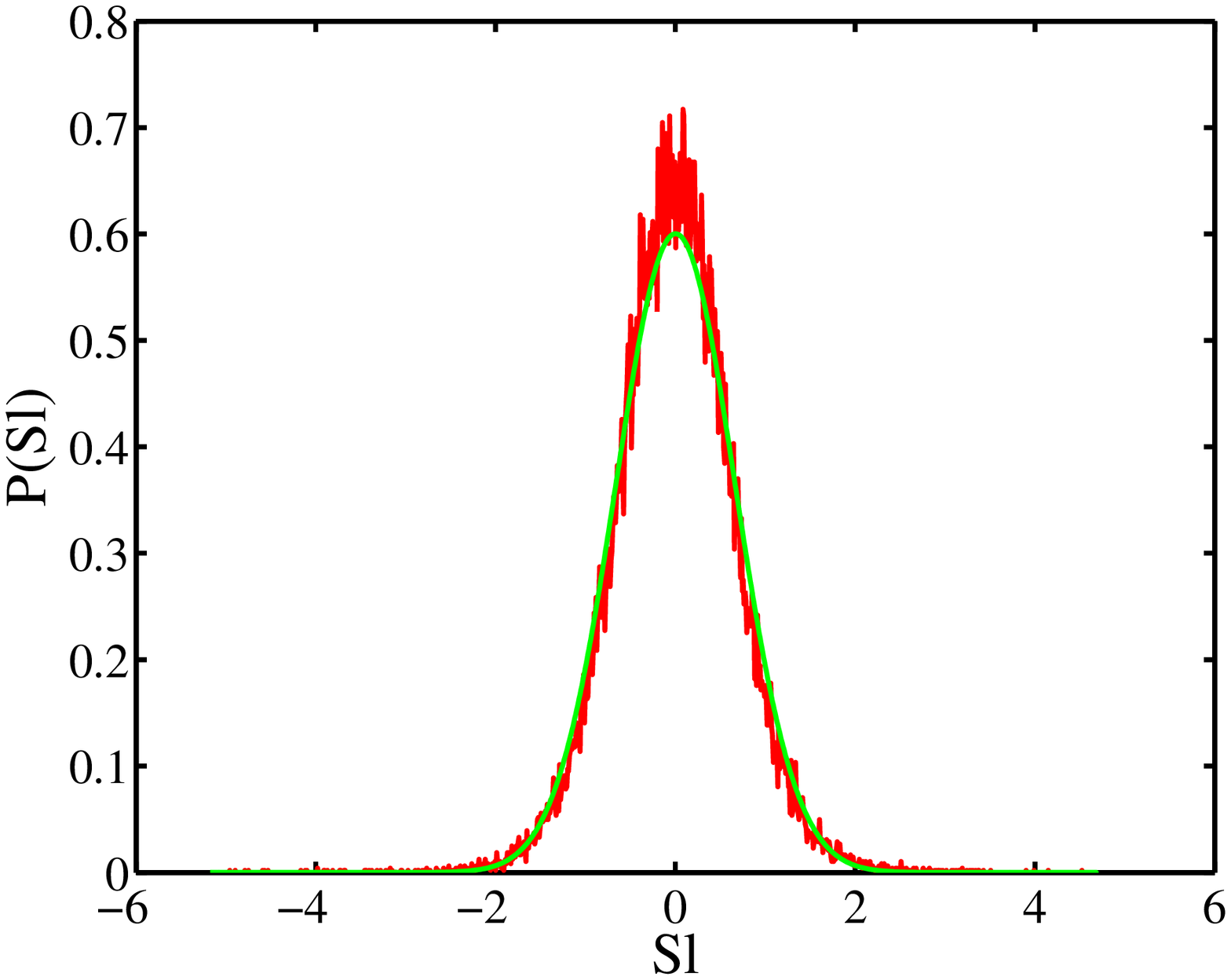}
    \includegraphics[angle=0,width=0.49\linewidth]{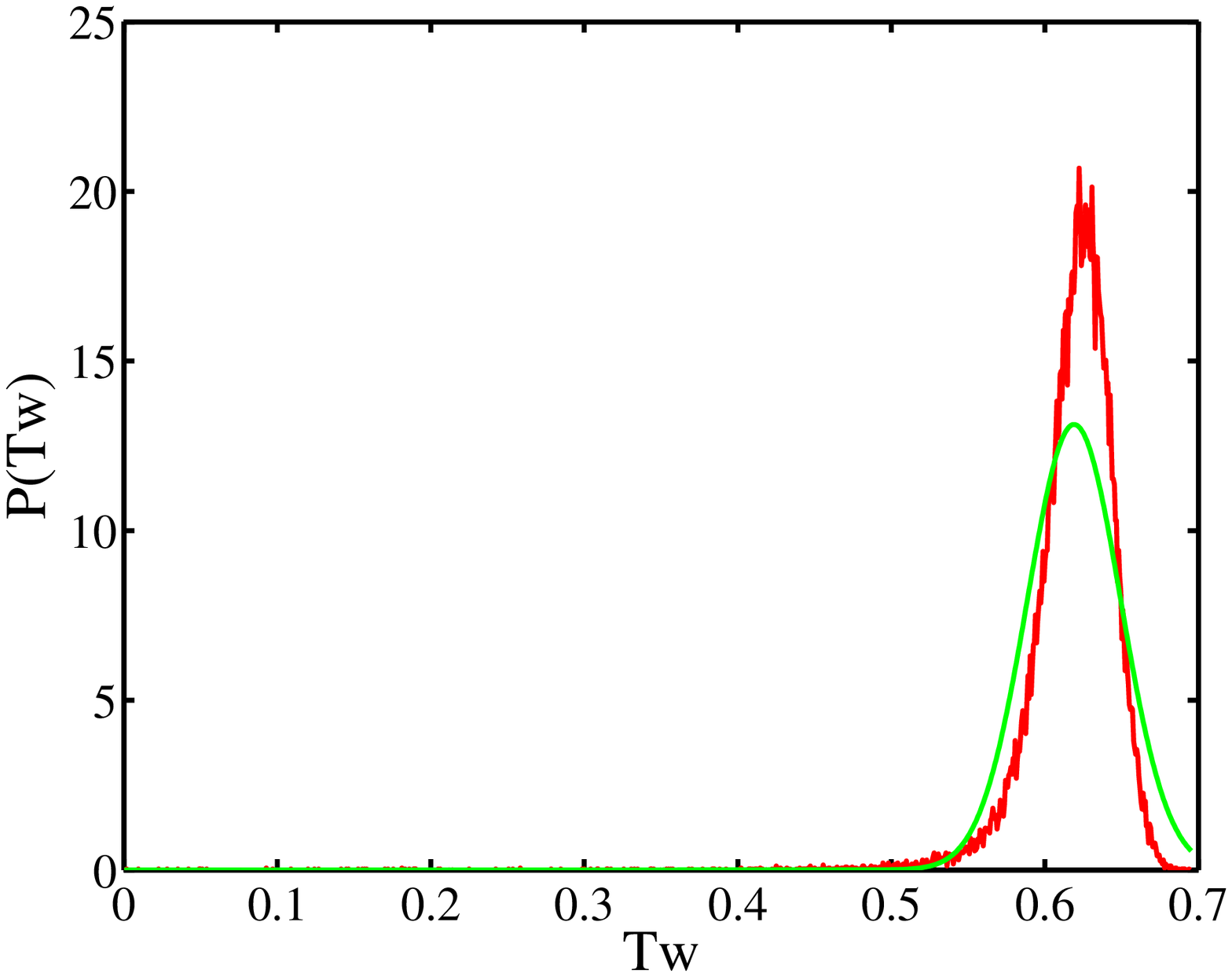}
    \includegraphics[angle=0,width=0.49\linewidth]{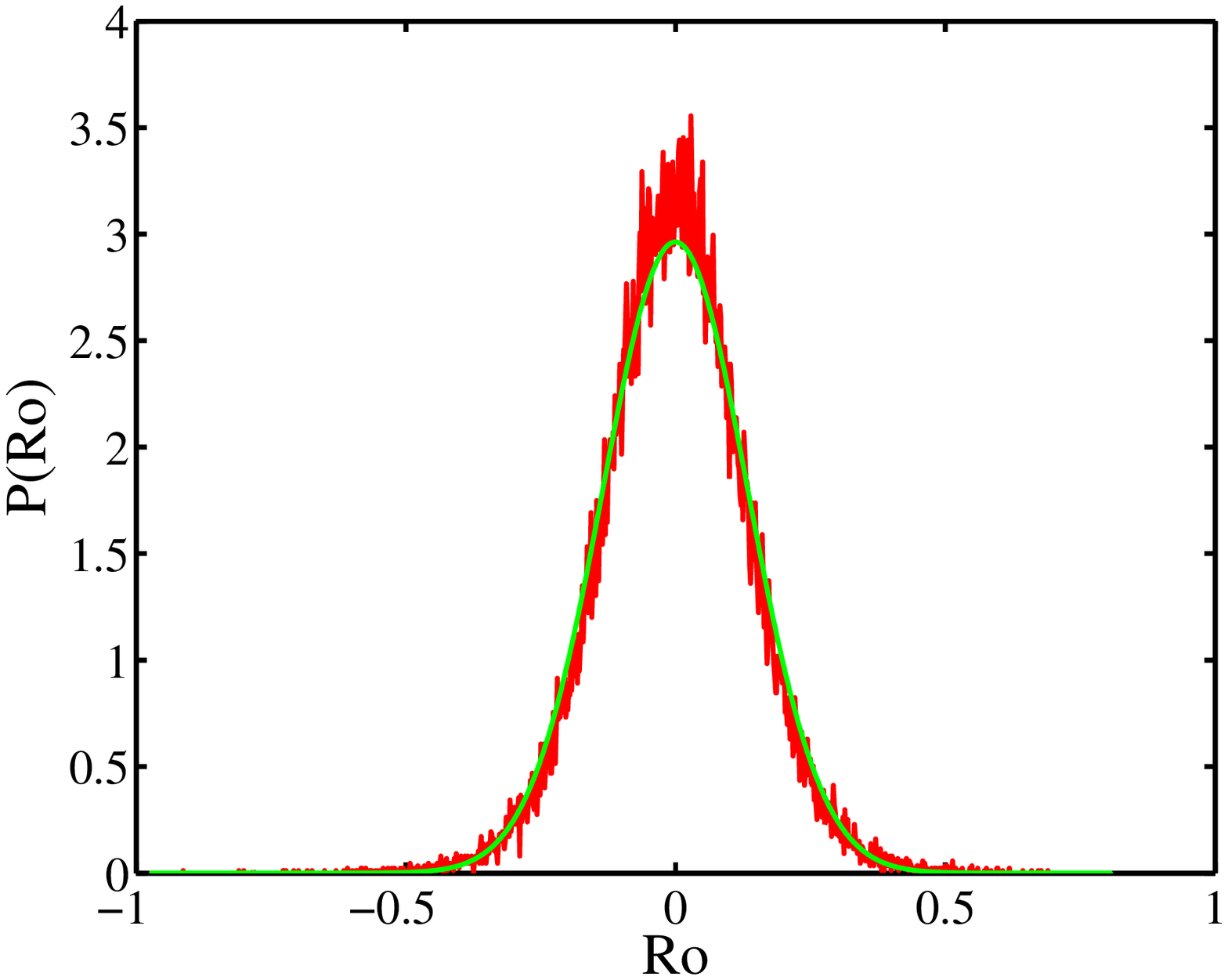}
    \includegraphics[angle=0,width=0.49\linewidth]{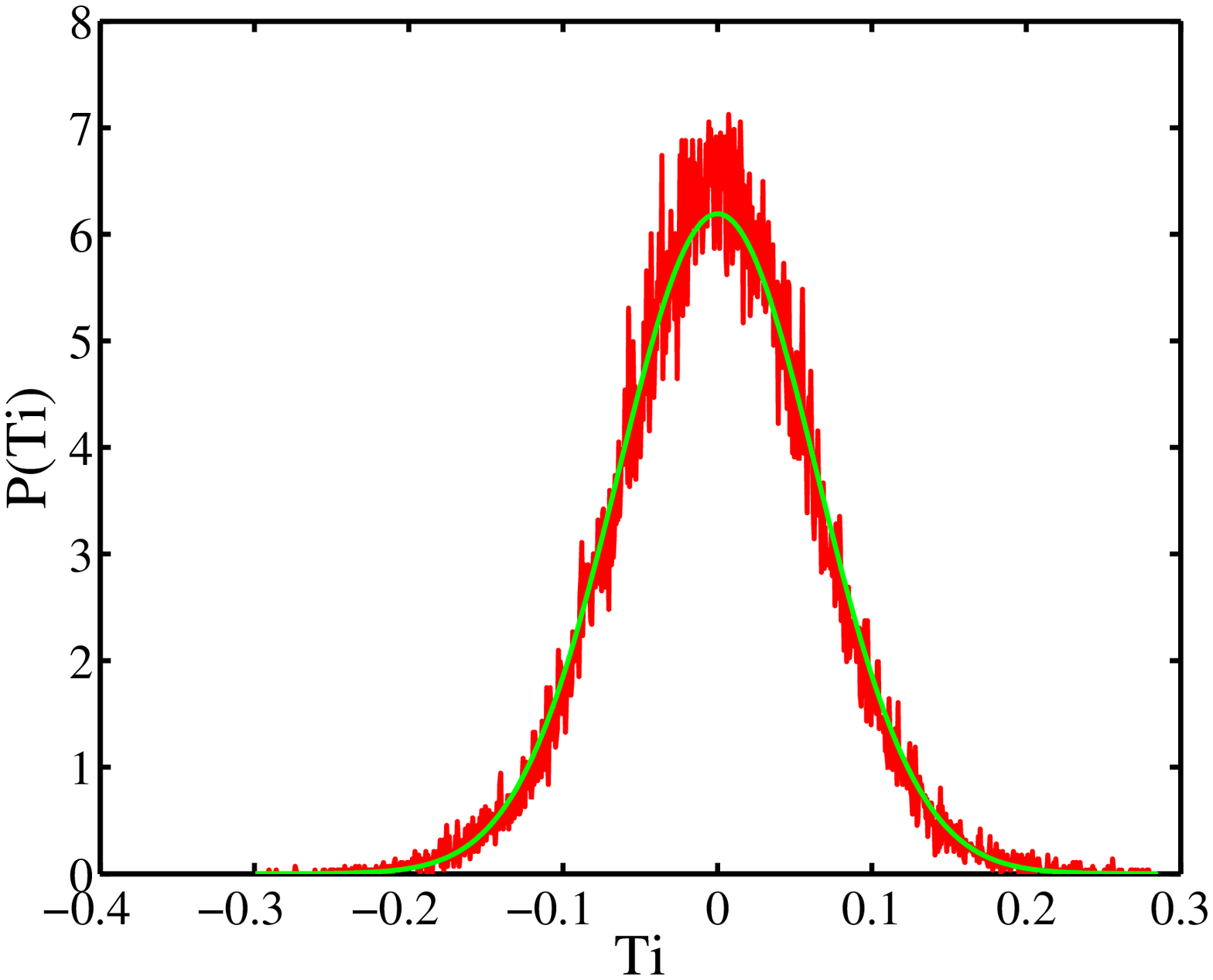}
    \end{center} \caption{(Color online) Comparison of probability distribution
      functions of all base-pair parameters for $\epsilon=20k_BT$,
      $k=64k_BT/\mbox{\AA}^2$, $2b=8\mbox{\AA}$. The Gaussians
      are plotted with the measured mean and mean squared values of
      the MC simulation.}
\label{fig:bpp}
\end{figure}
The calculation of the probability distribution functions of all six
base-pair parameters shows that especially the rise and twist motion
do not follow a Gaussian behavior. The deviation of the distribution
functions from the Gaussian shape depends mainly on the stacking
energy determined by $\epsilon$. For smaller values of $\epsilon$ one
observes larger deviations than for large $\epsilon$ values.

It is worthwhile to mention that there are mainly two correlations
between the base-pair parameters. The first is a microscopic
twist-stretch coupling determined by a correlation of Ri and Tw, i.e.
an untwisting of the helix implicates larger rise values. A
twist-stretch coupling was introduced in earlier rod models
\cite{Kamien_epl_97,Marko_epl_97,Nelson_bpj_98} motivated by
experiments with torsionally constrained DNA \cite{Strick_sci_96}
which allow for the determination of this constant. Here it is the
result of the preferred stacking of neighboring base-pairs and the
rigid backbones. The second correlation is due to constrained tilt
motion.  If we return to our geometrical ladder model we recognize
immediately that a tilt motion alone will always violate the
constraint of fixed backbone length $l$.  Even though we allow for
backbone fluctuations in the simulation the bonds are very rigid which
makes tilting energetically unfavorable. To circumvent this constraint
tilting always involves a directed shift motion.

Fig. \ref{fig:scat} shows that we recover the anisotropy of the
bending angles Ro and Ti as a result of the spatial dimensions of the
ellipsoids. Since the overlap of successive ellipsoids is larger in
case of rolling it is more favorable to roll than to tilt.
\begin{figure}[t]
  \begin{center}
    \includegraphics[angle=0,width=0.99\linewidth]{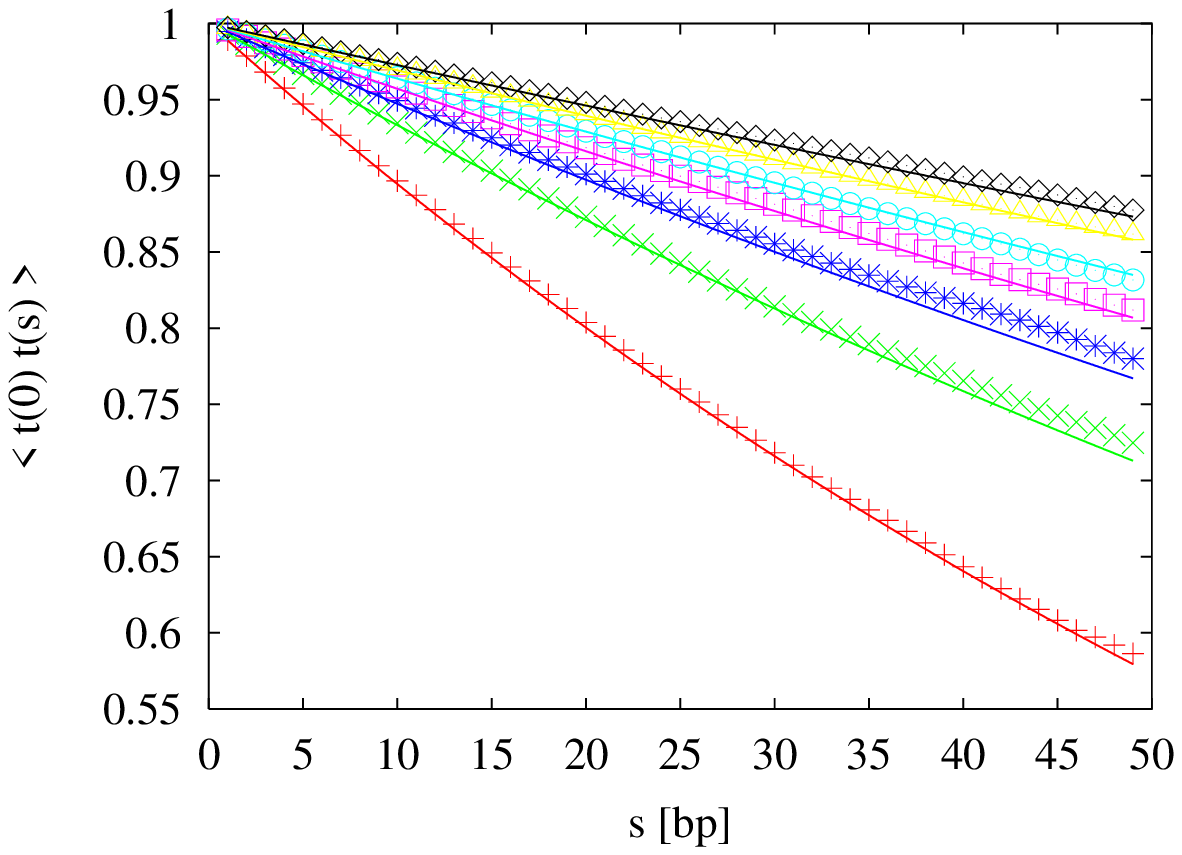}
    \includegraphics[angle=0,width=0.99\linewidth]{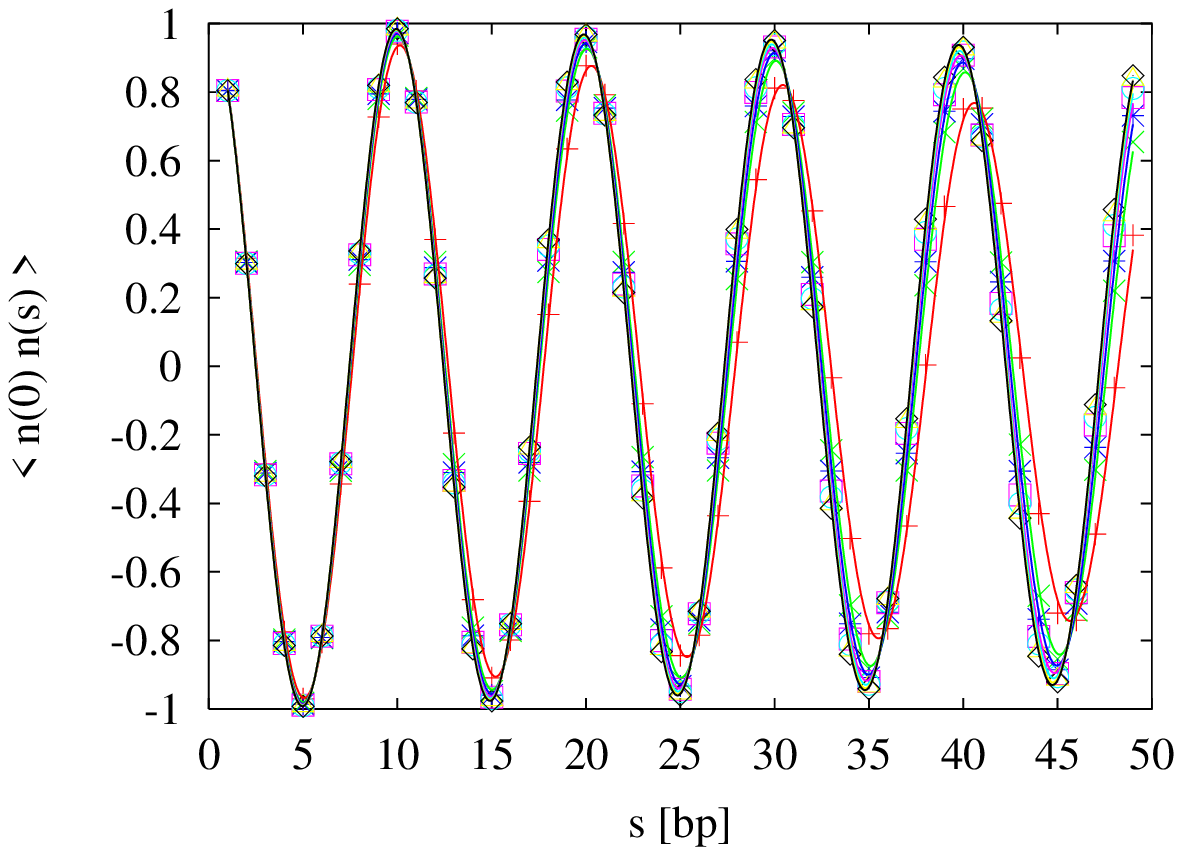}
    \end{center} \caption{(Color online) Comparison of analytical expressions
      Eqs.~(\ref{eq:lp}) and (\ref{eq:ln}) for $l_p$ and $l_n$ (solid
      lines) with numerically calculated orientational correlation
      functions (data points) for $2b=8\mbox{\AA}$,
      $k=64k_BT/\mbox{\AA}^2$, and $\epsilon=20,\,\ldots,\,60$
      [$k_BT$] (from bottom to top).}
\label{fig:lpln}
\end{figure}

The correlations can be quantified by calculating the correlation
matrix ${\cal C}$ of Eq.~(\ref{eq:sopcorr}). Inverting ${\cal C}$
yields the effective coupling constants of the SOP model ${\cal K} =
{\cal C}^{-1}$. Due to the local interactions it suffices to calculate
mean and mean squared values of Ri, Sl, Sh, Tw, Ro, and Ti
characterizing the 'internal' couplings of the base-pairs:
\begin{equation}
\label{eq:localcorrmat}
{\cal C} = (\sigma)_{ij},\,\forall i,j\in\{1,\ldots,6\}
\end{equation}
with $\sigma_{x,y}=\langle xy \rangle-\langle x \rangle \langle y
\rangle$.

\subsection{Bending and torsional rigidity}

\begin{figure}[t]
  \begin{center}
    \includegraphics[angle=0,width=0.99\linewidth]{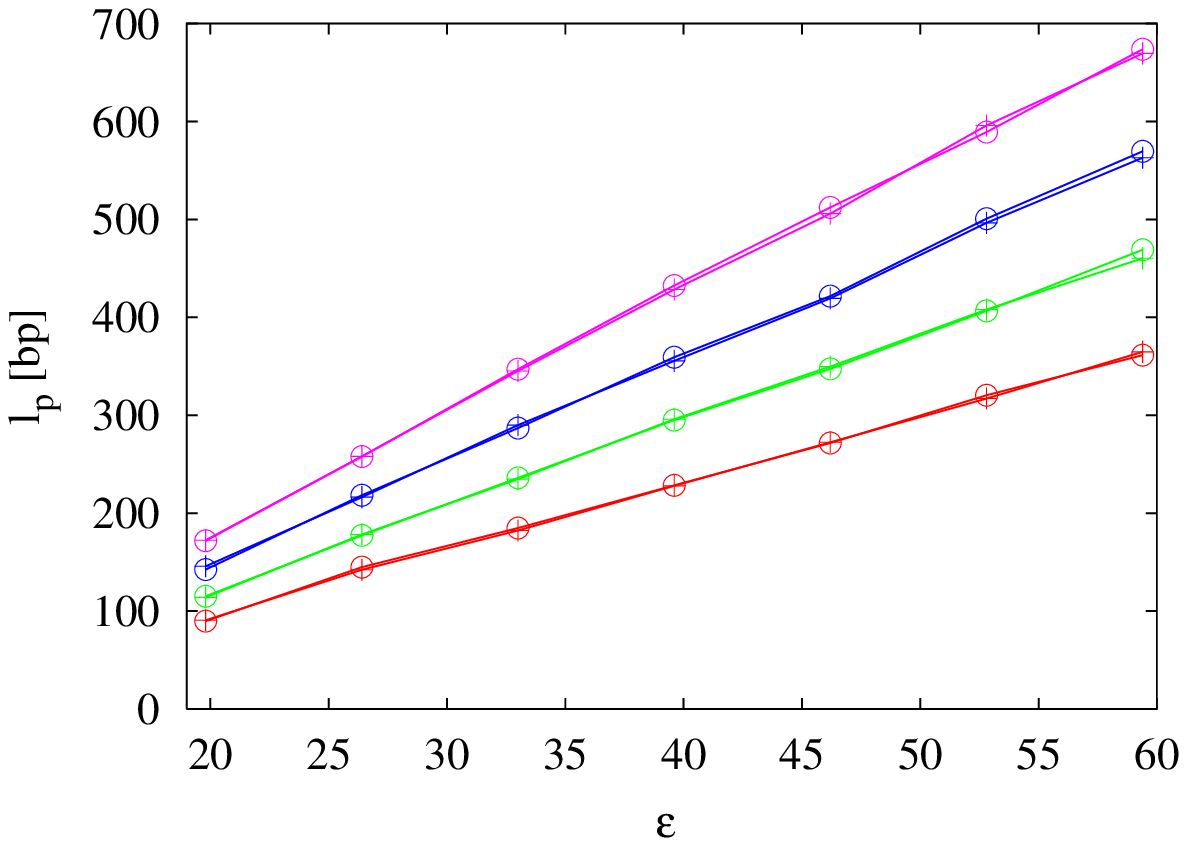}
    \includegraphics[angle=0,width=0.99\linewidth]{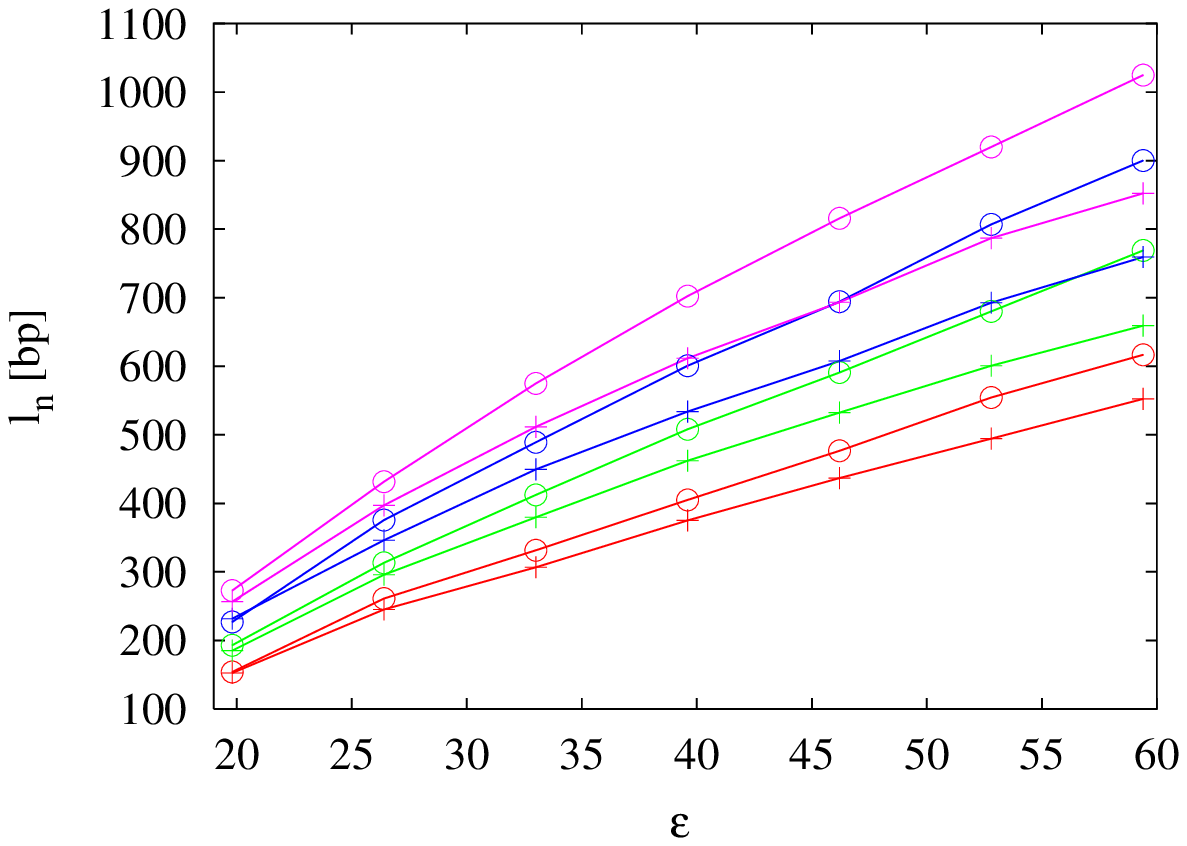}
  \end{center} \caption{(Color online) Dependency of (a) bending persistence length $l_p$
    and (b) torsional persistence length $l_n$ on the spring constant
    $k$, the width of the ellipsoids $b$ and the energy depth
    $\epsilon$. We measured the persistence lengths for varying width
    sizes $2b=8,\,9,\,10,\,11\mbox{\AA}$ (red, green, blue, purple)
    and for two different spring constants $k=32\mbox{
      (plus)},\,64\mbox{ (circles)}$ [$k_BT/\mbox{\AA}^2$].
    The bending persistence length depends solely on $b$ and
    $\epsilon$. It gets larger for larger $\epsilon$ and $b$ values.
    But it does not depend on $k$ (the curves for different $k$ values
    corresponding to the same width $b$ lie one upon the other). The
    torsional persistence length in turn depends on $k$, since a
    change of twist for constant Ri is proportional to a change in
    bond length.}
  \label{fig:lplnY}
\end{figure}
The correlation matrix of Eq.~(\ref{eq:localcorrmat}) can also be used
to check eqs.~(\ref{eq:lp}) and~(\ref{eq:ln}). Therefore we measured
the orientational correlation functions
$\langle\mathbf{t}_i\cdot\mathbf{t}_j\rangle$,
$\langle\mathbf{n}_i\cdot\mathbf{n}_j\rangle$,
$\langle\mathbf{b}_i\cdot\mathbf{b}_j\rangle$ and compared the results
to the analytical expressions as it is illustrated in
Fig. \ref{fig:lpln}. The agreement is excellent.

The simulation data show that the bending persistence length does not
depend on the spring constant $k$. But it strongly depends on
$\epsilon$ being responsible for the energy that must be paid to tilt
or roll two respective base pairs. Since a change of twist for
constant Ri is proportional to a change in bond length the bond energy
contributes to the twist persistence length explaining the dependence
of $l_{Tw}$ on $k$ (compare Fig. \ref{fig:lplnY}).

\begin{figure}[t]
  \begin{center}
    \includegraphics[angle=0,width=0.99\linewidth]{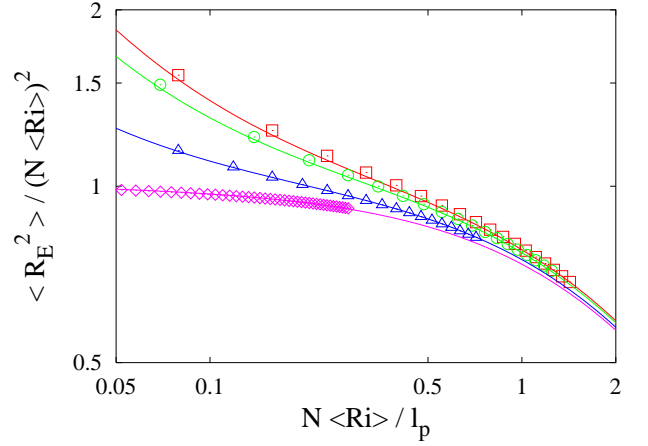}
    \end{center} \caption{(Color online) Comparison of the simulation data with
      $\epsilon=20k_BT$, $k=64k_BT/\mbox{\AA}^2$, $2b=11\mbox{\AA}$,
      and $T=1,2,3,5$ (from top to bottom) to Eqs.
      (\ref{eq:lp}), (\ref{eq:meanREstretchWLC}) and
      (\ref{eq:stretchmod}) (solid lines). Using the measured
      bending persistence lengths and the stretching moduli we find a
      good agreement with the predicted behavior. For $T=1$ we obtain
      $\gamma=6.02\mbox{\AA}^{-1}$.}
\label{fig:meanREstretchmod}
\end{figure}

We also measured the mean-square end-to-end distance $\langle
R_E^2\rangle$ and find that $\langle R_E^2\rangle$ deviates from the
usual WLC chain result due to the compressibility of the chain. So
as to investigate the origin of the compressibility we calculate
$\langle R_E^2\rangle$ for the following geometry. We consider two
base-pairs without spontaneous bending angles such that the end-to-end
vector $\vec{R}_E$ can be expressed as
\begin{equation}
  \vec{R}_E = \sum_i \vec{c}_i =  \sum_i ( \mbox{Ri}\, {\mathbf t}_i +
  \mbox{Sh}\, {\mathbf b}_i + \mbox{Sl}\, {\mathbf n}_i).
\end{equation}
The coordinate system $\{{\mathbf t}_i,{\mathbf b}_i,{\mathbf n}_i\}$
is illustrated in Fig. \ref{fig:bp}. $\vec{c}_i$ denotes the
center-center distance of two neighboring base-pairs. Since successive
base-pair step parameters are independent of each other, and
$\mbox{Ri}$ and $\mbox{Sh}$ and $\mbox{Sl}$ are uncorrelated the
mean-square end-to-end distance $\langle R_E^2\rangle$ is given by
\begin{equation}
  \label{eq:meanREstretchWLC}
  \begin{split}
    \langle R_E^2\rangle &= \sum_i(\langle
    c_i^2\rangle-\langle\mbox{Ri}\rangle^2) + \sum_i\sum_j
    \langle\mbox{Ri}\rangle^2 \langle{\mathbf t}_i\cdot{\mathbf
      t}_j\rangle\\
    &=
    \frac{N\langle\mbox{Ri}\rangle}{\gamma} + 2 N\langle\mbox{Ri}\rangle l_p-2l_p^2
    \left(
      1-\exp\left(-\frac{N\langle\mbox{Ri}\rangle}{l_p}\right)\right).
  \end{split}
\end{equation}
$N$ denotes the number of base-pairs. Note that
$\langle\mbox{Sl}\rangle$ and $\langle\mbox{Sh}\rangle$ vanish. Using
$\langle c_i^2\rangle = \langle\mbox{Ri}^2\rangle +
\langle\mbox{Sh}^2\rangle + \langle\mbox{Sl}^2\rangle$ the stretching
modulus $\gamma$ is simply given by
\begin{equation}
  \label{eq:stretchmod}
  \gamma = \frac{\langle\mbox{Ri}\rangle}
  {(\langle\mbox{Ri}^2\rangle-\langle\mbox{Ri}\rangle^2) +
    \langle\mbox{Sh}^2\rangle + \langle\mbox{Sl}^2\rangle}.
\end{equation}
We compared the data for different temperatures $T$ to Eq.
(\ref{eq:meanREstretchWLC}) using the measured bending persistence
lengths $l_p$ and stretching moduli $\gamma$ (see Fig.
\ref{fig:meanREstretchmod}). The agreement is excellent.
This indicates that {\em{transverse}} slide and shift fluctuations
contribute to the {\em{longitudinal}} stretching modulus of the chain.

\subsection{Stretching} \label{sec:str}

Extension experiments on double-stranded B-DNA have shown that the
overstretching transition occurs when the molecule is subjected to
stretching forces of $65\mbox{pN}$ or more~\cite{Smith_cosb_00}. The
DNA molecule thereby increases in length by a factor of $1.8$ times
the normal contour length. This overstretched DNA conformation is
called S-DNA. The structure of S-DNA is still under discussion. First
evidence of possible S-DNA conformations were provided by Lavery {\em
  et al.}~\cite{Cluzel_sci_96,Lavery_genetica_99,Lavery_jpcm_02} using
atomistic computer simulations.

In principle one can imagine two possible scenarios how the transition
from B-DNA to S-DNA occurs within our model. Either the chain untwists
and unstacks resulting in an untwisted ladder with approximately $1.8$
times the equilibrium length, or the chain untwists and the base-pairs
slide against each other resulting in a skewed ladder with the same
S-DNA length. The second scenario should be energetically favorable
since it provides a possibility to partially conserve the stacking of
successive base-pairs. In fact molecular modeling of the DNA
stretching
process~\cite{Cluzel_sci_96,Lavery_genetica_99,Lavery_jpcm_02} yielded
both a conformation with strong inclination of base-pairs and an
unwound ribbon depending on which strand one pulls.

We expect that the critical force $f_{crit}$ where the structural
transition from B-DNA to overstretched S-DNA occurs depends only on
the GB energy depth $\epsilon$ controlling the stacking energy. So as
a first step to find an appropriate value of $\epsilon$ as input
parameter for the MC simulation we minimize the Hamiltonian with an
additional stretching energy $E_{pull}=fc_{i,i+1}$, where the
stretching force acts along the center-of-mass axis, with respect to
$\mbox{Ri}$, $\mbox{Sl}$ and $\mbox{Tw}$ for a given pulling force
$f$.
\begin{figure}[t]
  \begin{center}
    \includegraphics[angle=0,width=0.99\linewidth]{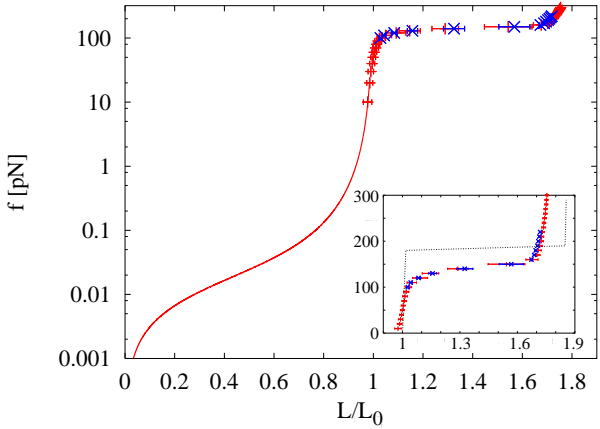}
    \end{center} \caption{(Color online) Force-extension relation calculated by minimum
      energy calculation (black) and obtained by MC simulation (red)
      for $50$ (red) and $500$ (blue) base-pairs. The red solid line
      represents the analytical result of the WLC. (inset) The
      deviation between energy minimization (black dotted line) and MC
      in the critical force is due to entropic contributions.}
    \label{fig:stst}
\end{figure}
Fig. \ref{fig:stst} shows the resulting stress-strain curve.  First
the pulling force acts solely against the stacking energy up to the
critical force where a jump from $L(f_{crit-})/L_0\approx1.05$ to
$L(f_{crit+})/L_0=\sqrt{\mbox{Ri}^2+\mbox{Sl}^2}/\mbox{Ri}\approx 1.8$
occurs, followed by another slow increase of the length caused by
overstretching the bonds. $L_0=L(F=0)=\mbox{Ri}$ denotes the
stress-free center-of-mass distance. As already mentioned three local
minima are obtained: (i) a stacked, twisted conformation, (ii) a
skewed ladder, and (iii) an unwound helix. The strength of the applied
stretching force determines which of the local minima becomes the
global one. The global minimum for small stretching forces is
determined to be the stacked, twisted conformation and the global
minima for stretching forces larger than $f_{crit}$ is found to be the
skewed ladder. Therefore the broadness of the force plateau depends
solely on the ratio of $l/\mbox{Ri}$ determined by the geometry of the
base pairs $S$ and the bond length $l=6.0\mbox{\AA}$. A linear
relationship is obtained between the critical force and the stacking
energy $\epsilon$ so that one can extrapolate to smaller $\epsilon$
values to extract the $\epsilon$ value that reproduces the
experimental value of $f_{crit}\approx65\mbox{pN}$. This suggests a
value of $\epsilon\approx7$.

\begin{figure}[t]
  \begin{center}
  \includegraphics[angle=0,width=0.90\linewidth]{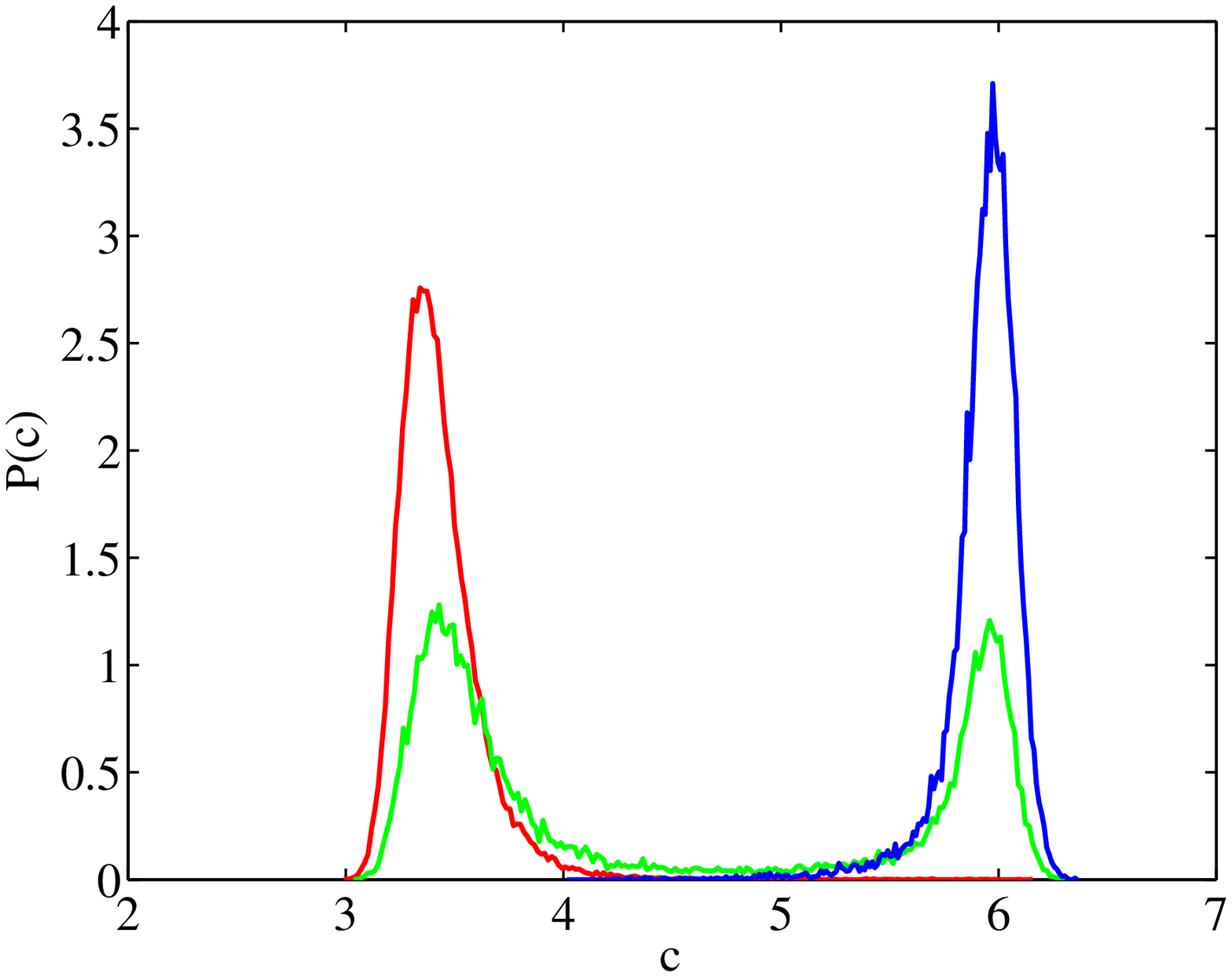}
  \includegraphics[angle=0,width=0.99\linewidth]{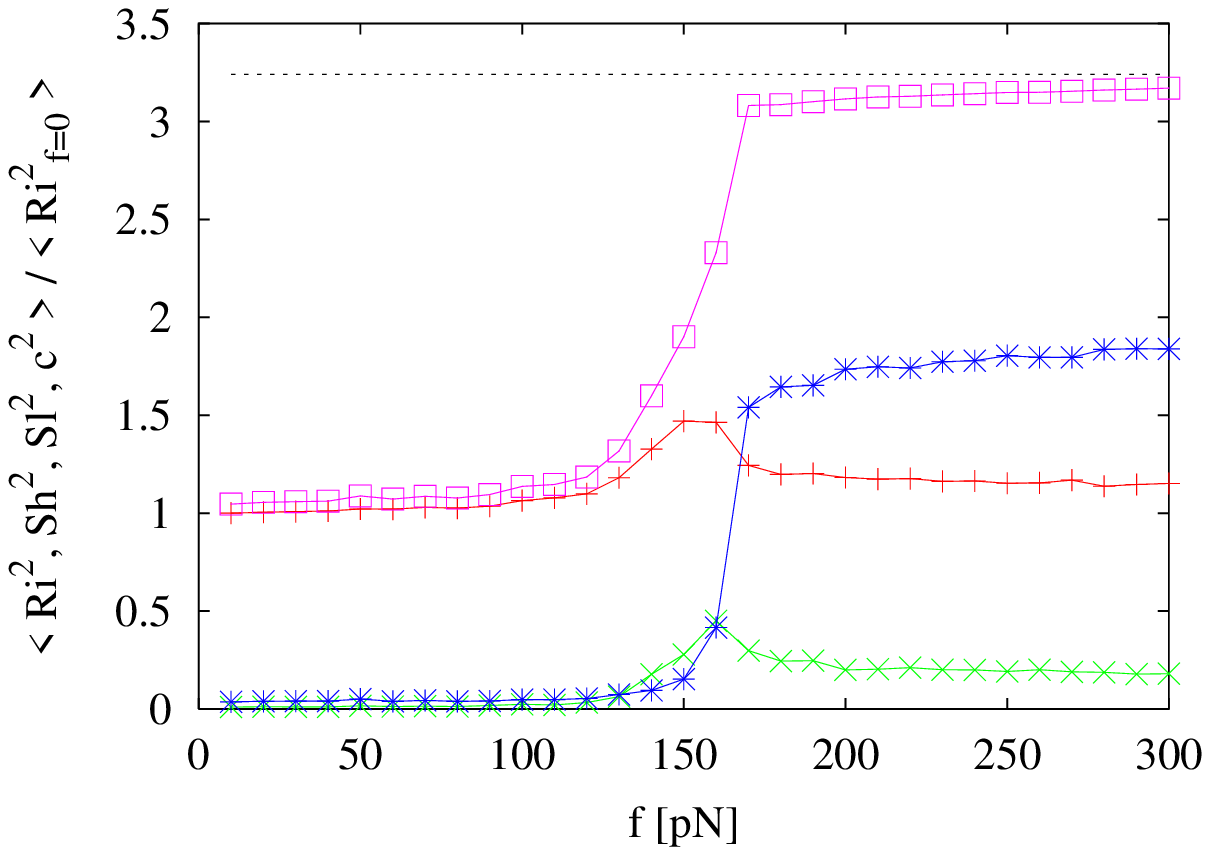}
  \end{center} \caption{(Color online) (a) Probability distribution function of the
    center-center distance of successive base-pairs for $f=0\mbox{
      (red)},\,140\mbox{ (green)},\,200\mbox{ (blue) pN}$. (b) Mean
    squared values of rise (red, plus), shift (green, crosses), slide
    (blue, stars), and center-of-mass distance (purple, squares) for
    neighboring base-pairs as a function of the stretching force $f$.
    The dashed line corresponds to the S-DNA center-of-mass distance.
    $\langle\mbox{Tw}\rangle$ of the resulting S-DNA conformation
    vanishes as predicted by Eq.~(\ref{twsl}).}
  \label{fig:re}
\end{figure}
The simulation results of the previous sections show several problems
when this value of $\epsilon$ is chosen. First of all it cannot
produce the correct persistence lengths, the chain is far to flexible.
Secondly the undistorted ground state is not a B-DNA anymore. The
thermal fluctuations suffice to unstack and untwist the chain locally.
That is why one has to choose larger $\epsilon$ values even though the
critical force is going to be overestimated.

Therefore we choose the following way to fix the parameter set
$\{b,\epsilon,k\}$. First of all we choose a value for the stacking
energy that reproduces correctly the persistence length. Afterwards
the torsional persistence length is fixed to the experimentally known
values by choosing an appropriate spring constant $k$. The depth of
the base-pairs has also an influence on the persistence lengths of the
chain. If the depth $b$ is decreased larger fluctuations for all three
rotational parameters are gained such that the persistence lengths get
smaller. Furthermore the geometric structure and the behavior under
pulling is very sensitive to $b$. Too small values provoke non-B-DNA
conformations or unphysical S-DNA conformations. We choose for $b$ a
value of $11\mbox{\AA}$ for those reasons. For $\epsilon=20$ and
$k=64$ a bending stiffness of $l_p=170\mbox{bp}$ and a torsional
stiffness of $l_n=270\mbox{bp}$ are obtained close to the experimental
values. We use this parameter set to simulate the corresponding
stress-strain relation.

The simulated stress-strain curves for $50$ base-pairs show three
different regimes (see Fig. \ref{fig:stst}). (i) For small stretching
forces the WLC behavior of the DNA in addition with linear stretching
elasticity of the backbones is recovered. This regime is completely
determined by the chain length $N$. Due to the coarse-graining
procedure that provides analytic expressions of the persistence
lengths depending on the base-pair parameters (see
eqs.~(\ref{eq:lp}),(\ref{eq:ln})) it is not necessary to simulate a
chain of a few thousand base-pairs. The stress-strain relation of the
entropic and WLC stretching regime (small relative extensions $L/L_0$
and small forces) is known
analytically~\cite{MarkoSiggia_mm_95,Odijk_mm_95}. Since we have
parameterized the model in such a way that we recover the elastic
properties of DNA on large length scales the simulation data for very
long chains will follow the analytical result for small stretching
forces. (ii) Around the critical force $f_{crit}\approx140\mbox{pN}$
which is mainly determined by the stacking energy of the base-pairs
the structural transition from B-DNA to S-DNA occurs.  (iii) For
larger forces the bonds become overstretched.  Our MC simulations
suggest a critical force $f_{crit}\approx140\mbox{pN}$ which is
slightly smaller than the value $f_{crit}\approx180\mbox{pN}$
calculated by minimizing the energy.  This is due to entropic
contributions.

In order to further characterize the B-to-S-transition we measured the
mean values of rise, slide, shift, etc. as a function of the applied
forces. The evaluation of the MC data shows that the mean values of
shift, roll and tilt are completely independent of the applied
stretching force and vanish for all $f$. Rise increases at the
critical force from the undisturbed value of $3.3\mbox{\AA}$ to
approximately $4.0\mbox{\AA}$ and decays subsequently to the
undisturbed value. Quite interestingly the mean value of slide jumps
from its undisturbed value of $0$ to $\pm5\mbox{\AA}$ (no direction is
favored) and the twist changes at the critical force from $\pi/10$ to
$0$. The calculation of the distribution function of the center-center
distance $c$ of two neighboring base-pairs for $f=140\mbox{pN}$ yields
a double-peaked distribution (see Fig. \ref{fig:re}) indicating that
part of the chain is in the B-form and part of the chain in the
S-form. The contribution of the three translational degrees of freedom
to the center-center distance $c$ is shown in Fig. \ref{fig:re}. The
S-DNA conformation is characterized by $\mbox{Ri}=3.3\mbox{\AA}$,
$\mbox{Sl}=\pm5\mbox{\AA}$ and $\mbox{Tw}=0$. In agreement with
Refs.~\cite{Cluzel_sci_96,Lavery_genetica_99} we obtain a conformation
with highly inclined base-pairs still allowing for partial stacking of
successive base-pairs.
\begin{figure}[t]
  \begin{center}
    \includegraphics[angle=0,width=0.85\linewidth]{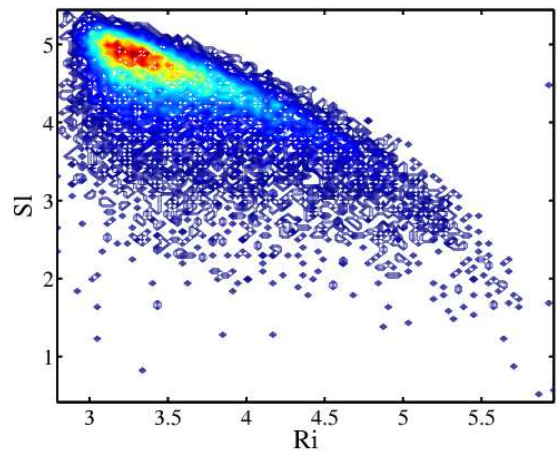}
    \end{center} \caption{(Color online) Contour plot of rise [$\mbox{\AA}$] versus
      slide [$\mbox{\AA}$] for the S-DNA conformation.}
    \label{fig:scatF20}
\end{figure}

\section{Discussion}

We have introduced a simple model Hamiltonian describing double-stranded DNA
on the base-pair level. Due to the simplification of the force-field and, in
particular, the possibility of non-local MC moves our model provides access to
much larger length scales than atomistic simulations. For example,
$4h$ on a AMD Athlon MP 2000+ processor are sufficient in order to 
generate 1000 independent conformations for chains consisting of
$N=100$ base-pairs. 

In the data analysis, the main emphasis was on deriving the elastic
constants on the elastic rod level from the analysis of thermal
fluctuations of base-pair step parameters. Assuming a twisted ladder
as ground state conformation one can provide an analytical
relationship between the persistence lengths and the local elastic
constants given by eqs. (\ref{eq:lp}), (\ref{eq:ln}) \footnote{The
  general case where the ground state is characterized by spontaneous
  rotations as well as spontaneous displacements as in the A-DNA
  conformation is more involved. This is the subject of ongoing
  work.}. Future work has to show, if it is possible to obtain
suitable parameters for our mesoscopic model from a corresponding
analysis of atomistic simulations~\cite{Lavery_bpj_00} or
quantum-chemical calculations~\cite{bickelhaupt}.  In the present
paper, we have chosen a top-down approach, i.e.  we try to reproduce
the experimentally measured behavior of DNA on length scales {\em
  beyond} the base diameter.  The analysis of the persistence lengths,
the mean and mean squared values of all six base-pair parameters and
the critical force, where the structural transition from B-DNA to
S-DNA takes place, as a function of the model parameters
$\{b,k,\epsilon\}$ and the applied stretching force $f$ suggests the
following parameter set:

\begin{align}
2b&=11\mbox{\AA}\\
\epsilon&=20k_BT\\
k&=64k_BT/\mbox{\AA}^2.
\end{align}

It reproduces the correct persistence lengths for B-DNA and entails
the correct mean values of the base-pair step parameters known by
X-ray diffraction studies. While the present model does not
include the distinction between the minor and major groove
and suppresses all  internal degrees of freedom of the base-pairs
such as propellor twist, it nevertheless reproduces some
experimentally observed features on the base-pair level.
For example, the anisotropy of the bending angles (rolling is
easier than tilting) is just a consequence of the plate-like shape of
the base-pairs and the twist-stretch coupling is the result of the
preferred stacking of neighboring base-pairs and the rigid backbones.

The measured critical force is overestimated by a factor of $2$ and
cannot be improved further by fine-tuning of the three free model
parameters $\{b,k,\epsilon\}$.  $f_{crit}$ depends solely on the
stacking energy value $\epsilon$ that cannot be reduced further.
Otherwise neither the correct equilibrium structure of B-DNA nor the
correct persistence lengths would be reproduced.  Our model suggests a
structure for S-DNA with highly inclined base-pairs so as to enable at
least partial base-pair stacking. This is in good agreement with
results of atomistic B-DNA simulations by Lavery {\em et
  al.}~\cite{Cluzel_sci_96,Lavery_genetica_99}. They found a force
plateau of $140\mbox{pN}$ for freely rotating
ends~\cite{Cluzel_sci_96}. The mapping to the SOP model yields the
following twist-stretch (Ri-Tw) coupling constant $k_{Ri,Tw} = ({\cal
  C}^{-1})_{Ri,Tw}=267/\mbox{\AA}$. $k_{Ri,Tw}$ is the microscopic
coupling of rise and twist describing the untwisting of the chain due
to an increase of rise (compare also Fig.  \ref{fig:scat}).

Possible applications of the present model include the investigation of
(i) the charge renormalization of the WLC elastic
constants~\cite{podgornik}, (ii) the microscopic origins of the
cooperativity of the B-to-S transition~\cite{Nelson_epl_03}, and (iii)
the influence of nicks in the sugar-phosphate backbone on
force-elongation curves. In particular, our model provides a
physically sensible framework to study the intercalation of certain
drugs or of ethidium bromide between base pairs.  The latter is a
hydrophobic molecule of roughly the same size as the base-pairs that
fluoresces green and likes to slip between two base-pairs forming an
DNA-ethidium-bromide complex. The fluorescence properties allow to
measure the persistence lengths of DNA~\cite{Schurr_arevpc_86}. It was
also used to argue that the force plateau is the result of a DNA
conformational transition~\cite{Cluzel_sci_96}.

In the future, we plan to generalize our approach to a description on
the base level which includes the possibility of hydrogen-bond
breaking between complementary bases along the lines of
Ref.~\cite{Barbi_pl_99,Cocco_prl_99}.  A suitably parameterized model
allows a more detailed investigation of DNA unzipping
experiments~\cite{Heslot_prl_97} as well as a direct comparison
between the two mechanism currently discussed for the B-to-S
transition: the formation of skewed ladder conformations (as in the
present paper) versus local
denaturation~\cite{Bloomfielda_bpj_01,Bloomfieldb_bpj_01,Bloomfieldc_bpj_01}.
Clearly, it is possible to study sequence-effects and even more
refined models of DNA. For example, it is possible to mimic minor and
major groove by bringing the backbones closer to one side of the
ellipsoids without observing non-B-DNA like ground states. The
relaxation of the internal degrees of freedom of the base-pairs
characterized by another set of parameters (propeller twist, stagger,
etc.) should help to reduce artifacts which are due to the ellipsoidal
shape of the base-pairs. Sequence effects enter via the strength of
the hydrogen bonds ($E_{GC}=2.9k_BT$ versus $E_{AT}=1.3k_BT$) as well
as via base dependent stacking interactions~\cite{Hunter_jmb_92}. For
example, one finds for guanine a concentration of negative charge on
the major-groove edge whereas for cytosine one finds a concentration
of positive charge on the major-groove edge. For adanine and thymine
instead there is no strong joint concentration of partial
charges~\cite{CalladineDrew99}.  It is known that in a solution of
water and ethanol where the hydrophobic effect is less dominant these
partial charges cause GG/CC steps to adopt A- or
C-forms~\cite{Fang_nar_99} by a negative slide and positive roll
motion and a positive slide motion respectively.  Thus by varying the
ratio of the strengths of the stacking versus the electrostatic energy
it should be possible to study the transition from B-DNA to A-DNA and
C-DNA respectively.

\section{Summary}

Inspired by the results of El Hassan and
Calladine~\cite{ElHassanCalladine_ptrs_97} and of Hunter et
al.~\cite{HunterLu_jmb_97,Hunter_jmb_92} we have put forward the idea
of constructing simplified DNA models on the base(-pair) level where
discotic ellipsoids (whose stacking interactions are modeled via
coarse-grained potentials~\cite{ralf_GB,Gay-Berne}) are linked to each
other in such a way as to preserve the DNA geometry, its major
mechanical degrees of freedom and the physical driving forces for the
structure formation~\cite{CalladineDrew99}.

In the present paper, we have used energy minimization and Monte Carlo
simulations to study a simple representative of this class of DNA
models with non-separable base-pairs. For a suitable choice of
parameters we obtained a B-DNA like ground state as well as realistic
values for the bend and twist persistence lengths. The latter were
obtained by analyzing the thermal fluctuations of long filaments as
well as by a systematic coarse-graining from the stack-of-plates to
the elastic rod level.  In studying the response of DNA to external
forces or torques, models of the present type are not restricted to
the regime of small local deformations. Rather by specifying a
physically motivated Hamiltonian for {\em arbitrary} base-(step)
parameters, our ansatz allows for realistic local structural
transitions. For the simple case of a stretching force we observed a
transition from a twisted helix to a skewed ladder conformation.
While our results suggest a similar structure for S-DNA as atomistic
simulations~\cite{Cluzel_sci_96}, the DNA model studied in this paper
can, of course, not be used to rule out the alternate possibility of
local strand
separations~\cite{Bloomfielda_bpj_01,Bloomfieldb_bpj_01,Bloomfieldc_bpj_01}.

In our opinion, the base(-pair) level provides a sensible compromise
between conceptual simplicity, computational cost and degree of
reality.  Besides providing access to much larger scales than
atomistic simulations, the derivation of such models from more
microscopic considerations provides considerable insight.  At the same
time, they may serve to validate and unify analytical approaches
aiming at (averaged) properties on larger
scales~\cite{ChatenayMarko_pre_01,HZhou_prl_99,Barbi_pl_99,Cocco_prl_99,Nelson_epl_03}.
Finally we note that the applicability of linked-ellipsoid models is
not restricted to the base-pair level of DNA as the same techniques
can, for example, also be used to study
chromatin~\cite{Wedemann_bpj_02,Katritch_jmb_00,mergell_prep}.

\section{Acknowledgments}

We greatfully acknowledge extended discussions with K. Kremer, R.
Lavery and A.C.  Maggs. We thank H. Schiessel for a careful reading of
our manuscript.  Furthermore we are greatful to the DFG for the
financial support of this work within the Emmy-Noether grant.

\bibliographystyle{phaip}

\begin{thebibliography}{10}

\bibitem{WatsonCrick_nat_53}
J.~D. Watson and F.~H.~C. Crick,
\newblock Nature {\bf 171}, 737 (1953).

\bibitem{Dickerson_sci_82}
R.~E. Dickerson et~al.,
\newblock Science {\bf 216}, 475 (1982).

\bibitem{DickersonFromAToZ}
R.~E. Dickerson,
\newblock Methods in Enzymology {\bf 211}, 67 (1992).

\bibitem{James_methenz_95}
T.~L. James,
\newblock Methods in Enzymology {\bf 261}, 1 (1995).

\bibitem{Millar_jcp_82}
D.~P. Millar, R.~J. Robbins, and A.~H. Zewail,
\newblock J. Chem. Phys. {\bf 76}, 2080 (1982).

\bibitem{Schurr_arevpc_86}
J.~M. Schurr and K.~S. Schmitz,
\newblock Annual Review of Physical Chemistry {\bf 37}, 271 (1986).

\bibitem{Perkins_sci_94}
T.~T. Perkins, S.~Quake, D.~Smith, and S.~Chu,
\newblock Science {\bf 264}, 822 (1994).

\bibitem{Boles_jmb_90}
T.~C. Boles, J.~H. White, and N.~R. Cozzarelli,
\newblock J. Mol. Biol. {\bf 213}, 931 (1990).

\bibitem{Smith_sci_92}
S.~B. Smith, L.~Finzi, and C.~Bustamante,
\newblock Science {\bf 258}, 1122 (1992).

\bibitem{Smith_sci_96}
S.~B. Smith, Y.~Cui, and C.~Bustamante,
\newblock Science {\bf 271}, 795 (1996).

\bibitem{Cluzel_sci_96}
P.~Cluzel et~al.,
\newblock Science {\bf 264}, 792 (1996).

\bibitem{Heslot_pnas_97}
B.~Essevaz-Roulet, U.~Bockelmann, and F.~Heslot,
\newblock Proc. Natl. Acad. Sci. USA {\bf 94}, 11935 (1997).

\bibitem{Allemand_pnas_98}
J.~Allemand, D.~Bensimon, R.~Lavery, and V.~Croquette,
\newblock Proc. Natl. Acad. Sci. USA {\bf 95}, 14152 (1998).

\bibitem{CalladineDrew_jmb_84}
C.~R. Calladine and H.~R. Drew,
\newblock J. Mol. Biol. {\bf 178}, 773 (1984).

\bibitem{Dickerson_emboj_89}
R.~E. Dickerson et~al.,
\newblock EMBO Journal {\bf 8}, 1 (1989).

\bibitem{LuOlson_jmb_99}
X.~J. Lu and W.~K. Olson,
\newblock J. Mol. Biol. {\bf 285}, 1563 (1999).

\bibitem{OlsonNomenclature_jmb_01}
W.~K. Olson et~al.,
\newblock J. Mol. Biol. {\bf 313}, 229 (2001).

\bibitem{CalladineDrew99}
C.~R. Calladine and H.~R. Drew,
\newblock {\em Understanding {DNA}: {T}he molecule and how it works},
\newblock Academic Press, 1999.

\bibitem{MarkoSiggia_mm_94}
J.~F. Marko and E.~D. Siggia,
\newblock Macromolecules {\bf 27}, 981 (1994).

\bibitem{MarkoSiggia_mm_95}
J.~F. Marko and E.~D. Siggia,
\newblock Macromolecules {\bf 28}, 8759 (1995).

\bibitem{Perkins_sci_95}
T.~T. Perkins, D.~E. Smith, R.~G. Larson, and S.~Chu,
\newblock Science {\bf 268}, 83 (1995).

\bibitem{Cozzarelli_90}
N.~R. Cozzarelli and J.~C. Wang,
\newblock {\em {{DNA} Topology and Its Biological Effects}},
\newblock Cold Spring Harbour Laboratory Press, Cold Spring Harbour, NY, 1990.

\bibitem{SchlickOlson_jmb_92}
T.~Schlick and W.~K. Olson,
\newblock J. Mol. Biol. {\bf 223}, 1089 (1992).

\bibitem{ChiricoLangowski_bp_94}
G.~Chirico and J.~Langowski,
\newblock Biopolymers {\bf 34}, 415 (1994).

\bibitem{Schiessel_prl_01}
H.~Schiessel, J.~Widom, R.~F. Bruinsma, and W.~M. Gelbart,
\newblock Phys. Rev. Lett. {\bf 86}, 4414 (2001).

\bibitem{ElHassanCalladine_londa_97}
M.~A.~E. Hassan and C.~R. Calladine,
\newblock Proc. R. Soc. Lond. A {\bf 453}, 365 (1997).

\bibitem{Hern_epjb_98}
C.~O'Hern, R.~Kamien, T.~Lubensky, and P.~Nelson,
\newblock Eur. Phys. J. B {\bf 1}, 95 (1998).

\bibitem{ChatenayMarko_pre_01}
A.Sarkar, J.~F. Leger, D.~Chatenay, and J.~F. Marko,
\newblock Phys. Rev. E {\bf 63}, 051903 (2001).

\bibitem{HZhou_prl_99}
Z.~Haijun, Z.~Yang, and O.-Y. Zhong-can,
\newblock Phys. Rev. Lett. {\bf 82}, 4560 (1999).

\bibitem{Barbi_pl_99}
M.~Barbi, S.~Cocco, and M.~Peyrard,
\newblock Physics Letters A {\bf 253}, 358 (1999).

\bibitem{Cocco_prl_99}
S.~Cocco and R.~Monasson,
\newblock Phys. Rev. Lett. {\bf 83}, 5178 (1999).

\bibitem{Lavery_Meso_bpj_99}
N.~Bruant, D.~Flatters, R.~Lavery, and D.~Genest,
\newblock Biophys. J. {\bf 77}, 2366 (1999).

\bibitem{ElHassanCalladine_ptrs_97}
M.~A.~E. Hassan and C.~R. Calladine,
\newblock Phil. Trans. R. Soc. Lond. A {\bf 355}, 43 (1997).

\bibitem{HunterLu_jmb_97}
C.~A. Hunter and X.-J. Lu,
\newblock J. Mol. Biol. {\bf 265}, 603 (1997).

\bibitem{Hunter_jmb_92}
C.~A. Hunter,
\newblock J. Mol. Biol. {\bf 230}, 1025 (1993).

\bibitem{ralf_GB}
R.~Everaers and M.~R. Ejtehadi,
\newblock Phys. Rev. E {\bf 67}, 041710 (2003).

\bibitem{Gay-Berne}
J.~G. Gay and B.~J. Berne,
\newblock J. Chem. Phys. {\bf 74}, 3316 (1981).

\bibitem{Olson_jmb_94b}
M.~S. Babcock, E.~P.~D. Pednault, and W.~K. Olson,
\newblock J. Mol. Biol. {\bf 237}, 125 (1994).

\bibitem{PackerHunter_jmb_98}
M.~J. Packer and C.~A. Hunter,
\newblock J. Mol. Biol. {\bf 280}, 407 (1998).

\bibitem{Binder_00}
D.~P. Landau and K.~Binder,
\newblock {\em Monte {C}arlo {S}imulations in {S}tatistical {P}hysics},
\newblock Cambridge University Press, 2000.

\bibitem{Metropolis_jcp_53}
N.~Metropolis, A.~W. Rosenbluth, M.~N. Rosenbluth, A.~N. Teller, and E.~Teller,
\newblock J. Chem. Phys. {\bf 21}, 1087 (1953).

\bibitem{Strick_genetica_99}
T.~R. Strick, D.~Bensimon, and V.~Croquette,
\newblock Genetica {\bf 106}, 57 (1999).

\bibitem{Lavery_genetica_99}
R.~Lavery and A.~Lebrun,
\newblock Genetica {\bf 106}, 75 (1999).

\bibitem{Lavery_jpcm_02}
R.~Lavery, A.~Lebrun, J.-F. Allemand, D.~Bensimon, and V.~Croquette,
\newblock J. Phys.: Condens. Matter {\bf 14}, R383 (2002).

\bibitem{Smith_cosb_00}
C.~Bustamante, S.~B. Smith, J.~Liphardt, and D.~Smith,
\newblock Current Opinion in Structural Biology {\bf 10}, 279 (2000).

\bibitem{Bensimon_epl_98}
D.~Bensimon, D.~Dohmi, and M.~Mezard,
\newblock Europhys. Lett. {\bf 42}, 97 (1998).

\bibitem{Bednar_jmb_95}
J.~Bednar et~al.,
\newblock J. Mol. Biol. {\bf 254}, 579 (1995).

\bibitem{Vologodskaia_jmb_02}
M.~Vologodskaia and A.~Vologodskii,
\newblock J. Mol. Biol. {\bf 317}, 205 (2002).

\bibitem{Kamien_epl_97}
R.~D. Kamien, T.~C. Lubensky, P.~Nelson, and C.~S. O'Hern,
\newblock Europhys. Lett. {\bf 38}, 237 (1997).

\bibitem{Marko_epl_97}
J.~F. Marko,
\newblock Europhys. Lett. {\bf 38}, 183 (1997).

\bibitem{Nelson_bpj_98}
P.~Nelson,
\newblock Biophys. J. {\bf 74}, 2501 (1998).

\bibitem{Strick_sci_96}
T.~R. Strick, J.-F. Allemand, D.~Bensimon, A.~Bensimon, and V.~Croquette,
\newblock Science {\bf 271}, 1835 (1996).

\bibitem{Odijk_mm_95}
T.~Odijk,
\newblock Macromolecules {\bf 28}, 7016 (1995).

\bibitem{Lavery_bpj_00}
I.~Lafontaine and R.~Lavery,
\newblock Biophys. J. {\bf 79}, 680 (2000).

\bibitem{bickelhaupt}
C.~F. Guerra and F.~M. Bickelhaupt,
\newblock Angewandte Chemie-International Edition {\bf 38}, 2942 (1999).

\bibitem{podgornik}
R.~Podgornik, P.~L. Hansen, and V.~A. Parsegian,
\newblock J. Chem. Phys. {\bf 113}, 9343 (2000).

\bibitem{Nelson_epl_03}
C.~Storm and P.~Nelson,
\newblock arXiv:physics/0212032  (2002).

\bibitem{Heslot_prl_97}
U.~Bockelmann, B.~Essevaz-Roulet, and F.~Heslot,
\newblock Phys. Rev. Lett. {\bf 79}, 4489 (1997).

\bibitem{Bloomfielda_bpj_01}
M.~C. Williams, J.~R. Wenner, I.~Rouzina, and V.~A. Bloomfield,
\newblock Biophys. J. {\bf 80}, 874 (2001).

\bibitem{Bloomfieldb_bpj_01}
I.~Rouzina and V.~A. Bloomfield,
\newblock Biophys. J. {\bf 80}, 882 (2001).

\bibitem{Bloomfieldc_bpj_01}
I.~Rouzina and V.~A. Bloomfield,
\newblock Biophys. J. {\bf 80}, 894 (2001).

\bibitem{Fang_nar_99}
Y.~Fang, T.~S. Spisz, and J.~H. Hoh,
\newblock Nucleic Acids Research {\bf 27}, 1943 (1999).

\bibitem{Wedemann_bpj_02}
G.~Wedemann and J.~Langowski,
\newblock Biophys. J. {\bf 82}, 2847 (2002).

\bibitem{Katritch_jmb_00}
V.~Katritch, C.~Bustamante, and W.~K. Olson,
\newblock J. Mol. Biol. {\bf 295} (2000).

\bibitem{mergell_prep}
B.~Mergell, H.~Schiessel, and R.~Everaers,
\newblock in preparation.

\end{thebibliography}

\end{document}